\documentclass[a4paper,fleqn]{cas-dc}
\usepackage{diagbox}
\usepackage{multirow}
\usepackage{lineno,hyperref}
\usepackage[numbers]{natbib}
\usepackage{subfigure}
\usepackage{amsfonts}
\usepackage{amssymb}
\usepackage{amsmath}
\usepackage{color,soul}
\usepackage{makecell}

\newtheorem{definition}{Definition}

\begin{document}

\shorttitle{}
\shortauthors{Mingyi Liu et al.}
\title [mode = title]{Data Correction and Evolution Analysis of the ProgrammableWeb Service Ecosystem} 
\author[1]{Mingyi Liu}

\address[1]{Faculty of Computing, Harbin Institute of Technology, Harbin, China}

\author[1]{Zhiying Tu}

\author[1]{Yeqi Zhu}

\author[1]{Xiaofei Xu}

\author[1]
{Zhongjie Wang}
\cormark[0]
\ead{rainy@hit.edu.cn}

\author[2]{Quan Z. Sheng}
\address[2]{Department of Computing, Macquarie University, Sydney, NSW 2109, Australia}

\cortext[cor0]{Corresponding author}

\begin{abstract}
The evolution analysis on Web service ecosystems has become a critical problem as the frequency of service changes on the Internet increases rapidly. 
Developers need to understand these evolution patterns to assist in their decision-making on service selection. \textit{ProgrammableWeb} is a popular Web service ecosystem on which several evolution analyses have been conducted in the literature. However, the existing studies have ignored the quality issues of the \textit{ProgrammableWeb} dataset and the issue of service obsolescence. In this study, we first report the quality issues identified in the \textit{ProgrammableWeb} dataset from our empirical study. Then, we propose a novel method to correct the relevant evolution analysis data by estimating the life cycle of application programming interfaces (APIs) and mashups. We also reveal how to use three different dynamic network models in the service ecosystem evolution analysis based on the corrected \textit{ProgrammableWeb} dataset. Our experimental experience iterates the quality issues of the original \textit{ProgrammableWeb} and highlights several research opportunities.
\end{abstract}


\maketitle

\section{Introduction}
With the development of Web 2.0 and the wide adoption of service-oriented architecture (SOA), many services now expose their features in the form of application programming interfaces (APIs). Multiple APIs can be easily composed into an application, also called \textit{mashups}~\cite{NguCSP10}, that creates and delivers unique new value to customers. This growing phenomenon is called the {\em API economy} \cite{Brown2014}. As domain barriers have opened and cross-border cooperation and cross-border integration have become more common, the promotion of the API economy in the Internet era has gradually weakened the concept of the traditional domain. These represent substantial changes to the traditional Internet ecosystem, leading to the evolution of service ecosystems.

The API economy has enabled businesses in regard to cross-border integration and innovation, 
creating an increasing number of new applications. Moreover, as users experience these new applications, they may discover new needs, which not only further accelerates the innovation process but also intensifies market competition. 
Service providers need to be sensitive to the changes of user needs and preferences and constantly bring forth new services. Therefore, studying the evolution of service ecosystems is important because it can offer insights and significant benefits from different perspectives. From a business perspective, evolution analysis assists decision-making by helping service providers and market regulators understand the evolutionary patterns of a service ecosystem, thereby guiding sustainable and healthy service/service ecosystem development. For example, through evolution analysis, service providers can learn the collaboration strategies of their competitors and discover popular market evolution trends, allowing them to adjust their business strategies for fleeting innovation opportunities and thereby maintaining or enhancing the competitiveness of their services. From a technical perspective, evolution analysis can mine interpretable prior knowledge from data to facilitate other downstream tasks, such as service recommendation, service discovery, and service composition, thereby accelerating the pace of service development~\cite{BouguettayaSHSD17,ShengQVSBX14}.

\textit{ProgrammableWeb}\footnote{https://www.programmableweb.com/.} is the largest online API store platform, and it collects a large number of third-party APIs and mashups. Every day, new APIs/mashups emerge, existing APIs perish, and different APIs dynamically cooperate to create new mashups. Thus, it is a typical Web service ecosystem. In addition, \textit{ProgrammableWeb} serves as a standard research dataset in the field of service computing. As a typical representative of the real Internet ecosystem, it has been employed to support many service science studies, especially in the fields of service recommendation~\cite{ma2020deep,botangen2020geographic,kalai2018social}, service discovery~\cite{adeleye2019fitness,XuLWSYW17}, service evolution analysis~\cite{tian2017exploratory}, and quality of service (QoS) prediction~\cite{chen2020accurate,chen2017exploiting}. As of January 9, 2021, \textit{ProgrammableWeb} collected $23,881$ APIs and $7,973$ mashups, with detailed information such as date of creation, category, profile, and active status.

\textbf{Motivation}: A number of studies have focused on the evolution analyses of the \textit{ProgrammableWeb} service ecosystem from a variety of perspectives. The analysis results of all the existing studies show a positive and optimistic attitude towards the healthy development of the \textit{ProgrammableWeb} service ecosystem. Some studies have also provided a generative model of the service ecosystem network to guide service recommendation and service discovery. These studies \cite{wittern2014graph, zhong2014time, huang2013impact,pan2018structure} are usually explicitly or implicitly based on two assumptions: i) the \textit{ProgrammableWeb} dataset is high quality and credible, and ii) the obsolescence of APIs and mashups and the impact of such demise can be ignored. However, these assumptions require more careful data processing because the ProgrammableWeb dataset has certain flaws. For example, quality issues in labeling, especially regarding the active availability status and obsolescence times of APIs and mashups, are important. Ignoring the demise of APIs and mashups makes it more difficult to assess the true health status of the service ecosystem, promotes blindly optimistic conclusions, leads to inaccurate evolutionary patterns, and damages downstream tasks.

\begin{figure}
\centering
\includegraphics[width=0.9\linewidth]{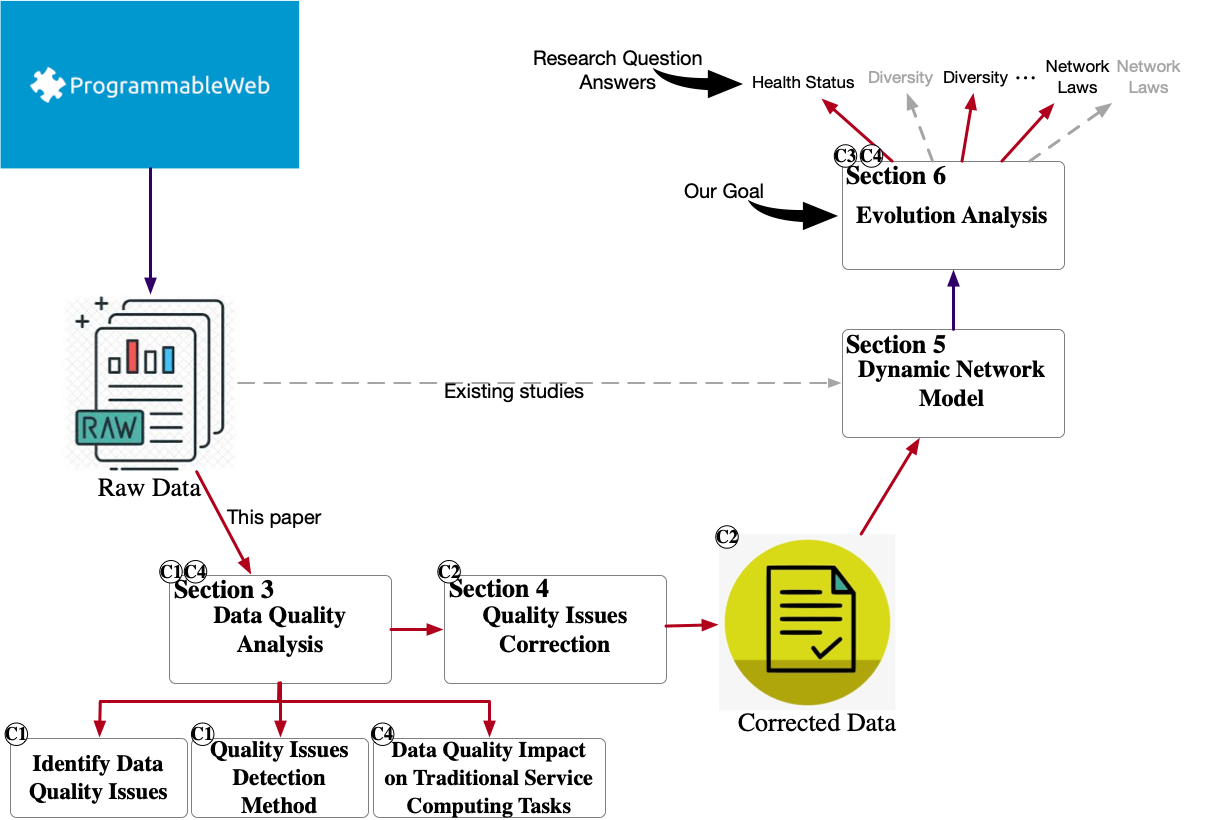}
\caption{Overview of this paper relative to existing studies in the context of evolution analysis.}
\label{fig:overview}
\end{figure}

Figure~\ref{fig:overview} shows the overview of this paper relative to existing studies in the context of evolution analysis. Existing studies directly use the raw data crawled from the \textit{ProgrammableWeb} for dynamic network modeling followed by evolutionary analysis. However, in this paper, we first identify and correct the erroneous information related to evolution in the \textit{ProgrammableWeb} dataset, specifically for the active state and time-related attributes. Then, we discuss how to use dynamic networks to model the service ecosystem. After establishing a dynamic network model, we analyze the evolution of the service ecosystem using the network properties and network visualization approaches. Finally, we discuss the problems that exist in the service ecosystem and their impacts on other service tasks.

The main contributions and innovations of this study are as follows:

\begin{itemize}
\item[C1] We identify data quality issues, including data incompleteness, data errors and data noise, in the commonly used \textit{ProgrammableWeb} dataset by exploiting statistical methods, automated network request testing, and manual inspections.
\item[C2] We correct the active status of APIs and mashups in the \textit{ProgrammableWeb} dataset based on the results of automated network request testing and propose a normal distribution-based method to estimate the active time of APIs and mashups. We have released the new \textit{ProgrammableWeb} dataset with the active status and active time corrected for other researchers to conduct related studies\footnote{The dataset is available on GitHub: https://github.com/HIT-ICES/Correted-ProgrammableWeb-dataset}. 
\item[C3] We conduct an evolution analysis based on the corrected \textit{ProgrammableWeb} dataset, and the results show that the original dataset is not suitable for service ecosystem research. The assumption of perfect data quality misleads statistical results of the data and affects the choice of algorithms. If these factors are not considered, the results of the existing research works based on the old dataset should be revisited, such as various conclusions about the current health status, diversity, and network laws of the \textit{ProgrammableWeb} service ecosystem.

\item[C4] During the evolution analysis, we analyze the potential impact of the data quality issues of the original \textit{ProgrammableWeb} dataset from both business and technical viewpoint, and provide some suggestions for addressing or avoiding these quality problems. We also highlight some opportunities for future research using the new \textit{ProgrammableWeb} dataset.
\end{itemize}

The remainder of this paper is organized as follows. Section \ref{sec:related_work} 
discusses the related works. Section \ref{sec:problems} presents various data quality issues related to the evolutionary analysis of the \textit{ProgrammableWeb} dataset. Section \ref{sec:corrected} provides a data restoration method based on probability estimation. Section \ref{sec:model} clarifies how to use dynamic networks to model service ecosystems. Section \ref{sec:RQ} explains the methods, reports the results of the evolution analysis of the service ecosystem, and discusses the impacts of other quality issues 
of the original \textit{ProgrammableWeb} dataset on traditional service computing tasks and the new challenges and opportunities that the corrected \textit{ProgrammableWeb} dataset introduces for traditional tasks. Finally, 
Section~\ref{sec:conclusion} offers some concluding remarks.

\section{Related Works}\label{sec:related_work}
In recent years, the evolution analysis of service ecosystems, such as the \textit{ProgrammableWeb}, has been studied extensively. The existing studies focus on individual service state changes and the changes to the service network topology that are intended to help developers select and integrate appropriate services into their own applications. The ultimate goal of these studies is to provide prior knowledge to help solve various traditional service computing problems, such as service recommendation and service label prediction~\cite{Yao-TSC2015}.

For example, Wang et al. \cite{wang2009mining} discovered user behavior patterns in mashup communities by studying the network and clustering properties of the \textit{ProgrammableWeb} service ecosystems. The authors argued that mashup communities possess scale-free properties and frequently used APIs attract large numbers of users. Weiss et al. \cite{weiss2010modeling} examined the structure of the mashup ecosystem and its growth over time and concluded that i) the distribution of mashups over APIs follows a power law and that 
ii) the complexity of mashups continues to increase. 
Huang et al. \cite{huang2014recommendation} proposed a three-phase network prediction approach (NPA) to study both usage patterns and the evolution traces of the entire \textit{Programm-ableWeb} service ecosystem for evolution-aware service recommendation. The authors in \cite{zhong2014time} considered the evolution of service ecosystems for service recommendation, extracted service evolution patterns by exploiting latent Dirichlet allocation (LDA) and time series predictions, and then used these patterns to guide service recommendation. Lyu et al. \cite{lyu2014three} proposed a three-level view model for Web service ecosystem visualization. Such visualization is valuable not only for understanding the ecosystem but also for providing support to service consumers, helping them to discover appropriate services. Bai et al. \cite{bai2017sr} developed a tailored topic model to mine effective representations of service ecosystems that addresses both service evolution and information sparsity, while Gan et al. \cite{gan2019multi} proposed a novel approach for service multilabel recommendation using deep neural networks.

\begin{figure*}
\centering
\includegraphics[width=\textwidth]{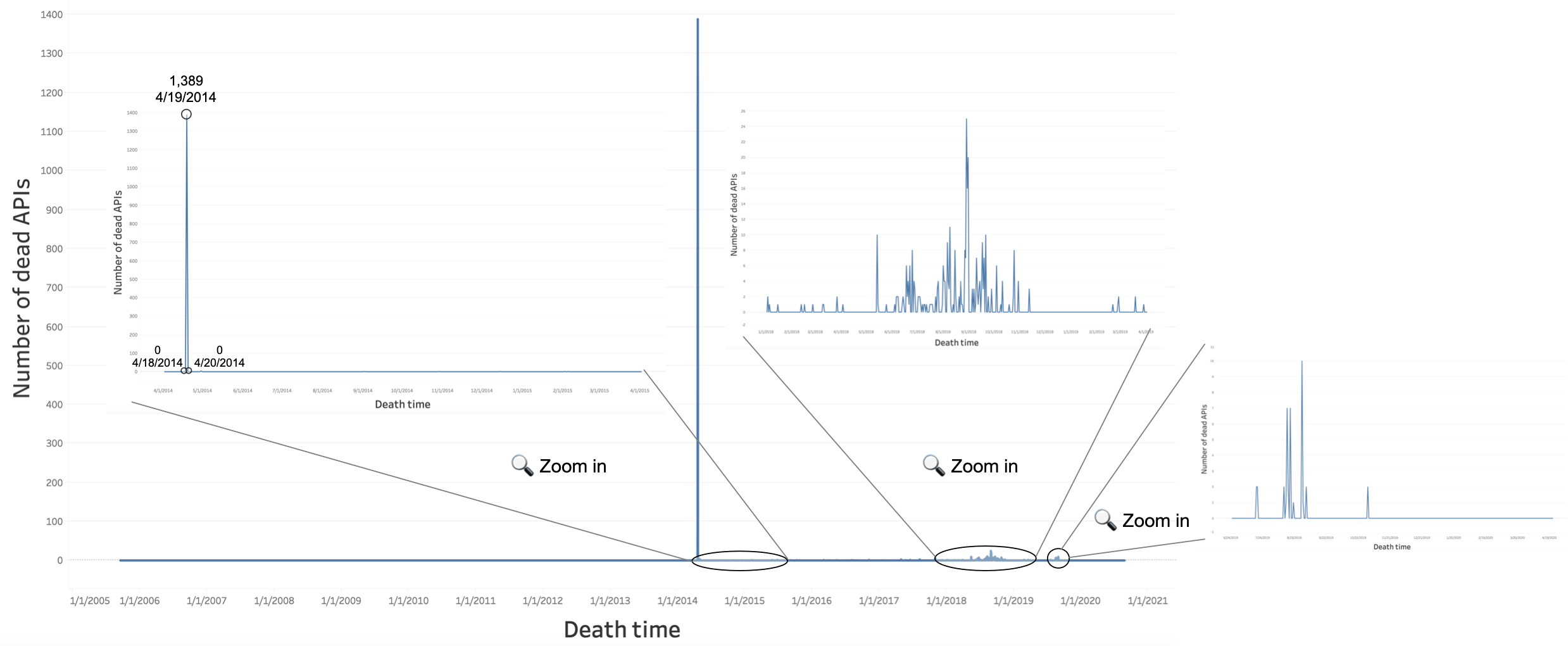}
\caption{The death time distribution of the APIs in the deathpool.}\label{fig:deathpoo_api}
\end{figure*}

In addition to exploring the Web service ecosystem represented by the \textit{ProgrammableWeb}, the evolution of other types of service ecosystems have also been explored in recent studies. For instance, the work in \cite{sampaio2017supporting} focuses on microservice ecosystems. Aggregating structural, deployment, and run-time information of an evolving microservice system into one model provides actionable insights to help developers manage service upgrades, architectural evolution, and changing deployment trade-offs. Wang et al. \cite{wang2018dkem} also focused on microservice ecosystems by proposing a distributed knowledge-based evolution model (DKEM) that can discover stable evolution patterns and automatically explore new and more stable cooperation among services. In addition to these technical-level service ecosystems, some researchers have turned their interests to the evolution of business-level service ecosystems. A very recent work in \cite{liu2020data,9283677} proposed a novel multilayer network-based service ecosystem model (MSEM) that can be constructed automatically by mining massive textual datasets from the Internet. Via the introduction of the concept of service events, MSEM can not only explore evolutionary patterns but also determine the driving factors of the evolution.

\section{Data Quality Issues}
\label{sec:problems}
We collected a total of $23,678$ APIs and $7,766$ mashups from the \textit{ProgrammableWeb} website (including data in the deathpool). After checking and testing these data, we identified three serious data quality issues related to prior evolutionary analyses using the original \textit{ProgrammableWeb} dataset. These issues are described in detail in the rest of this section.

\subsection{Untrustworthy Death Time}
\textit{ProgrammableWeb} places the deprecated APIs and mash\-ups into a collection called the ``deathpool''. The time at which a deprecated API or mashup enters the deathpool is marked and this time is often referred to as the ``death time`` of the API or the mashup.

\begin{figure*}
\centering
\includegraphics[width=0.8\textwidth]{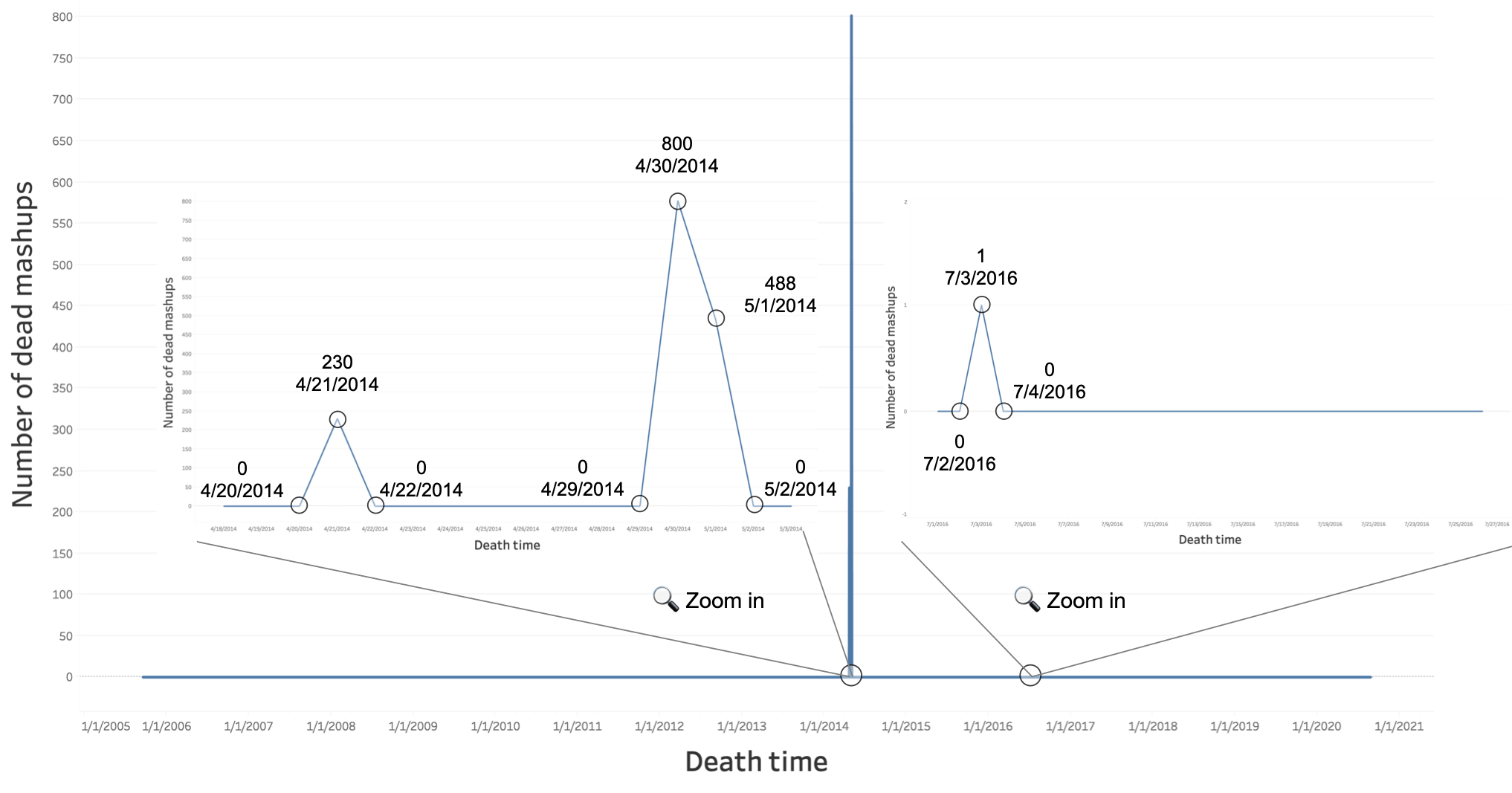}
\caption{The death time distribution of the mashups in the deathpool.}\label{fig:deathpoo_mashup}
\end{figure*}

We conducted statistical studies on the death times of APIs and mashups in the deathpool provided by \textit{Programma\-bleWeb} and the results are shown in Figure~\ref{fig:deathpoo_api} and Figure~\ref{fig:deathpoo_mashup}. The results indicate that the death times of most APIs and mashups in the deathpool are clustered at specific periods (e.g., April 2014). Interestingly, some APIs/mashups are created even later than the creation time; for example, \textit{Iron Mountain Policy Center}\footnote{https://www.programmableweb.com/api/iron-mountain-policy-center} was marked as deprecated in April 2014 but was actually submitted on January 21, 2020. This might be the result of the automatic processing at a certain time because the death time labels are obviously unreasonable and are therefore unsuitable for Web service evolution studies. In terms of distribution, the distribution of death times between 2018 and 2020 is more in line with the prior knowledge of experts. We also compiled statistics on the survival duration of APIs that died during this period, as shown in Figure~\ref{fig:survival_days}, and found no data in the results that violate common sense, such as a service that dies before its birth.
To further confirm the reliability of this portion of the data, we randomly selected $20$ APIs and investigated their real death times. We found that the average death time error was $50$ days, which is an acceptable result. Therefore, we believe that recent data (those with death times from 2018 to 2020) are reliable estimations of API/mashup death times. This aspect will be discussed in more detail in Section~\ref{sec:estimate}.

\subsection{Incorrect API Status} 
We tested the status of APIs that was marked as available in two ways:
\begin{enumerate}
\item We detected whether the descriptive text of the API explicitly indicates its available status. When the text indicates ``no longer available``, we consider the API's status to be dead (or obsolete).
\item For the rest of the APIs,
we tested the results of accessing their API endpoints (URL addresses) under strict conditions to prevent available APIs from being marked as dead APIs when it is determined that the API is not available. We conclude that an API is regarded as obsolete only when its URL is not reachable or when the network access request returns a 404 status code.
\end{enumerate}

After using automatic network request tests to check\footnote{All the network request tests were conducted on the Google Cloud platform. We repeated the tests three times on different dates and in different periods and manually reviewed 100 randomly selected results.} the availability status of all APIs and manually analyzing their descriptive texts, we summarize three different death patterns of unavailable APIs, as illustrated in Figure~\ref{fig:api_pattern}:

\begin{figure*}
\centering
\includegraphics[width=\textwidth]{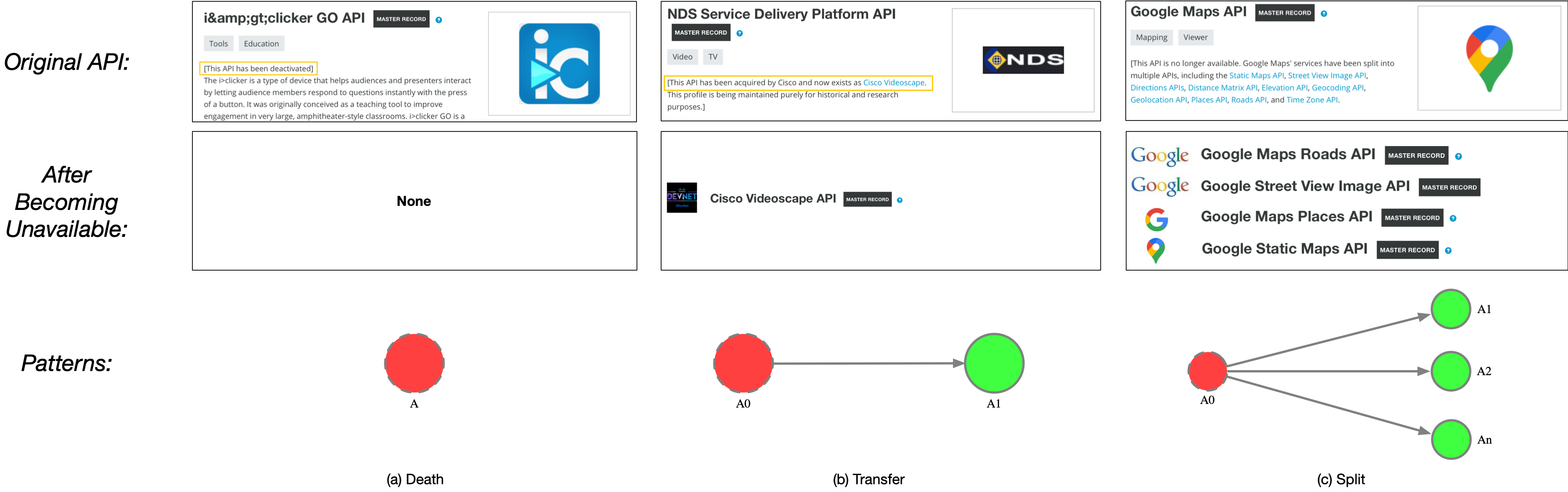}
\caption{Three different API death patterns: the red circles represent unavailable APIs, while the green circles represent available APIs.}
\label{fig:api_pattern}
\end{figure*}

\begin{itemize}
\item \textit{Death}: The API no longer provides its services in any form; i.e., it has entirely and permanently perished from the ecosystem.
\item \textit{Transfer}: The original API is no longer available, but all its functions were transferred to another API that continues to provide services. This pattern usually occurs when an API is acquired or the name is changed.
\item \textit{Split}: The original API is no longer available; its services have been split into several independent APIs, and each independent API provides a portion of the original API's services.
\end{itemize}

Table~\ref{tab:api_pattern} shows the statistical results of the available status of the APIs on \textit{ProgrammableWeb}. 
From the table, we can see that only approximately $44.7\%$ of the APIs are truly active and that only $14.8\%$ of the unavailable APIs are correctly marked as unavailable (including APIs in the deathpool). In addition, $85.2\%$ of the unavailable APIs are incorrectly marked as active. The proportions of different death patterns are listed in Table~\ref{tab:api_pattern}. Most of the unavailable APIs (over $99.9\%$) are caused by the death of the APIs. The number of unavailable APIs caused by \textit{transfer} and \textit{split} is relatively small, $17$ and $7$, respectively. However, the APIs that are unavailable due to \textit{split} and \textit{transfer} have a substantial impact on the API service ecosystem. For example, the most commonly used APIs, such as \textit{Google Map}, \textit{Facebook}, and \textit{Twitter}, have all split. The statistical results in Figure~\ref{fig:mashup_api_death} also support this point.

\begin{table}[!bhtp]
\centering
\caption{The statistical results of the available status of the APIs on \textit{ProgrammableWeb. All the APIs in the deathpool are marked as \textbf{dead}.}* denotes the overall percentage, ** denotes the row percentage.}\label{tab:api_pattern}

\begin{tabular}{l|l|rr} 
\hline
\multicolumn{2}{l|}{\diagbox{Truth}{Labeled}} & Available & Unavailable \\ 
\hline
\multicolumn{2}{l|}{Available} & \makecell[c]{10,534 \\ ($44.7\%^*$)} & - \\ 
\hline
\multirow{3}{*}{Unavailable} & Dead & 11,087 & \multirow{3}{*}{\makecell[c]{1,934\\($14.8\%^{**}$)}} \\
& Split & 7 & \\
& Transfer & 17 & \\
\hline
\end{tabular}
\end{table}

\subsection{Incorrect Mashup Information}

\begin{figure*}
\centering
\includegraphics[width=0.9\textwidth]{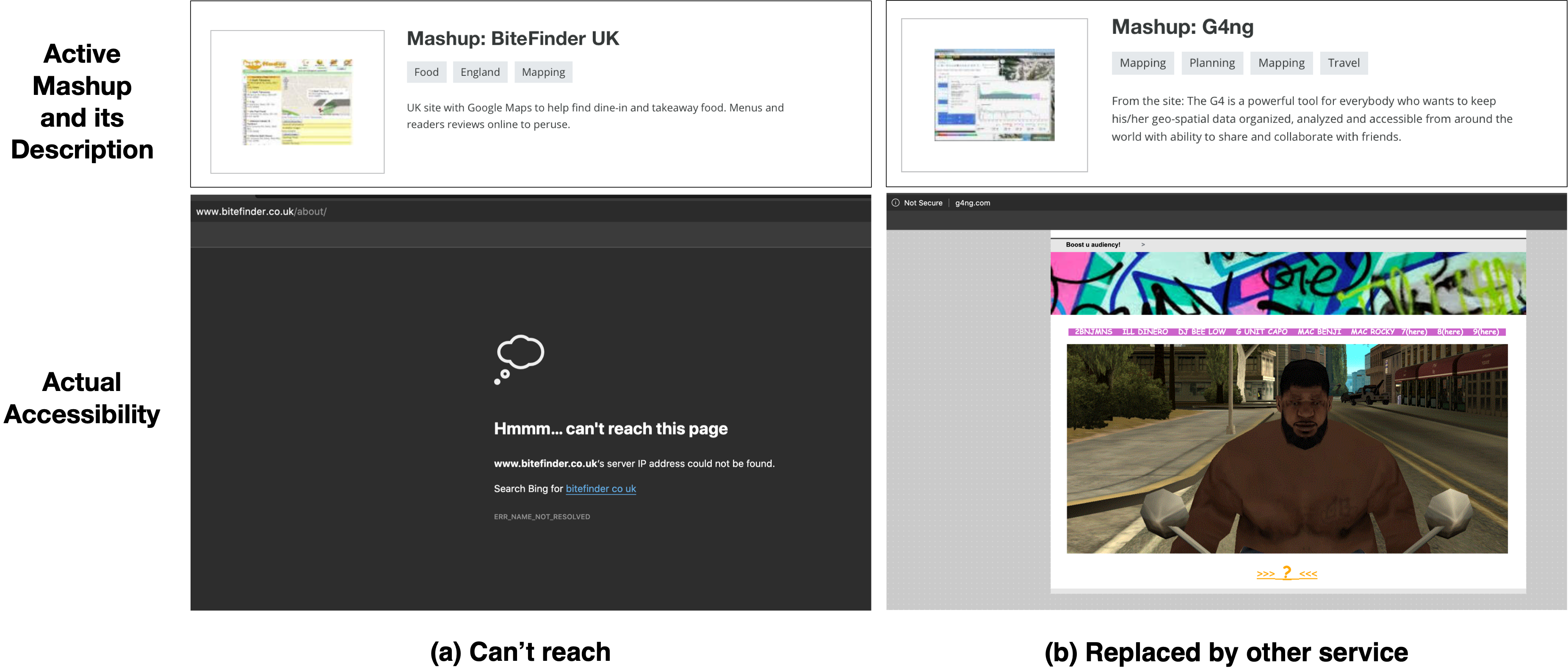}
\caption{Two different mashup death patterns.} \label{fig:mashup_death_pattern}
\end{figure*}

The data quality issues in mashups are primarily twofold: a) {\em incorrect activity status} and b) {\em incorrect API composition information}.

An incorrect activity status means that mashups cannot properly provide the services they describe but are still marked as available.
We invited 5 coders who are all graduate students working on service computing topics, to perform manual analysis. For each mashup, two coders are randomly selected for manual checking. Coders were required to report whether content in the homepage of this mashup is inconsistent with the description in the page of this mashup on \textit{ProgrammableWeb}. For those mashups that did not match, they were asked to check the history information at WebArchive\footnote{http://web.archive.org} and then summarized the reasons for the unavailability. When two coders come to different conclusions on one mashup, this mashup will be voted by the remaining three coders. Finally, we use Inter-Rater Reliability (IRR)[28] to measure the level of agreement. During the manual checking, we did not find any cases that different coders have different conclusions on one mashup, and this is mainly because of the relevant professional background of all five coders and the clear criteria for their judgment. Therefore, the IRR is 100\% agreement between coders on 100\% samples, which is sufficient agreement among multiple coders. Based on the provided report, we conclude that mashups are unavailable due to two different patterns: \textit{unreachable} and \textit{replaced}. Figure~\ref{fig:mashup_death_pattern} illustrates these two patterns with examples. Based on the above two patterns, the remaining mashups are determined to be unavailable if they satisfy any of the following conditions:
\begin{itemize}
\item \textit{Unreachable}: The homepage is not accessible or does not exist (404).
\item \textit{Replaced}: The homepage's HTML source code does not contain information about the mashup.
\end{itemize}

We manually checked 530 mashups and this number is greater than 366 (the minimum number of mashups for statistical representation with a 95\% confidence level and 5 intervals). The 530 mashups were selected based on the following steps:
\begin{itemize}
    \item[Step 1]330 mashups were manually checked at random, including 130 available mashups and 200 unavailable mashups. The purpose of this step is to build mashup status detection algorithm. \textbf{We stopped when we checked 200 unavailable mashups because 1) for these unavailable mashups, coders not only need to determine their availability, but more importantly, to identify exact unavailable time of these mashups by searching on http://web.archive.org, which is an extremely time-consuming task. 2) during the examination, we found that after 50 unavailable mashups were found, no new reasons for mashup unavailability emerged, so we believe that 200 unavailable mashups are sufficient for summarising the patterns of mashups unavailability and providing sufficient a priori knowledge for the algorithm to be constructed.}
    \item[Step 2] 200 mashups were randomly selected from mashups marked as \textit{unavailable} by the algorithm for manual check. The purpose of this step is to check the validity of the rule-based algorithm we have constructed. We chose 200 to be consistent with the number of 200 unavailable mashups in the previous step.
\end{itemize}

The test results show that the true numbers of available and unavailable mashups are 2,489 (32.0\%) and 5,277 (68.0\%), respectively, while the numbers given by \textit{ProgrammableWeb} are 6,247 (80.4\%) and 1,519 (19.6\%), respectively (including mashups in the deathpool).

\begin{figure}[htbp]
\centering
\subfigure[Ratio of all mashups that invoke different types of unavailable APIs]{
\includegraphics[width=0.85\linewidth]{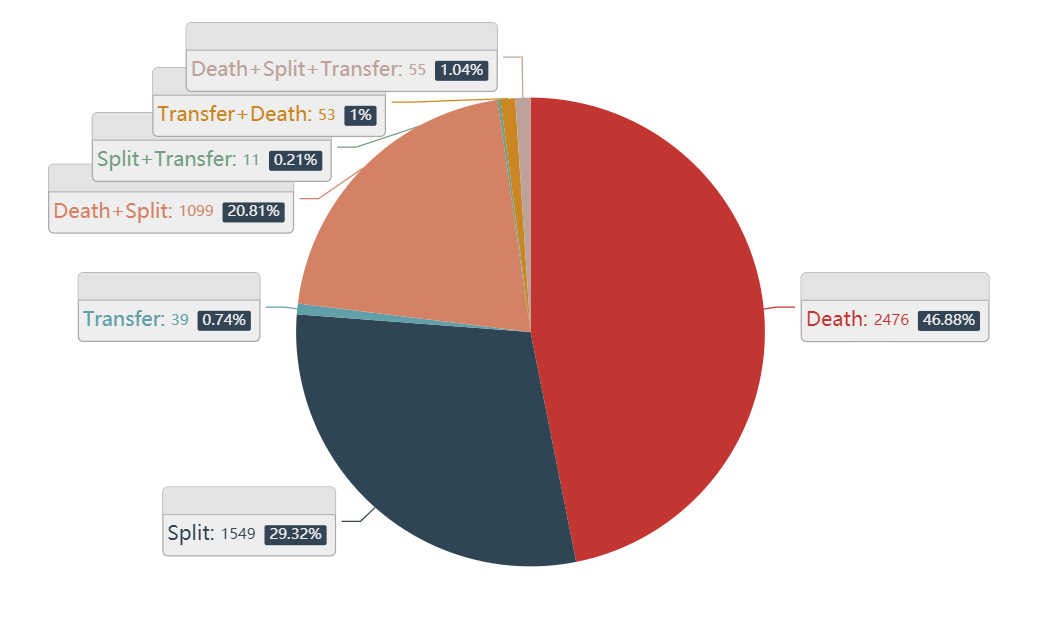}\label{fig:mashup_api_death_a}
}
\subfigure[Ratio of available mashups that invoke different types of unavailable APIs]{
\includegraphics[width=0.85\linewidth]{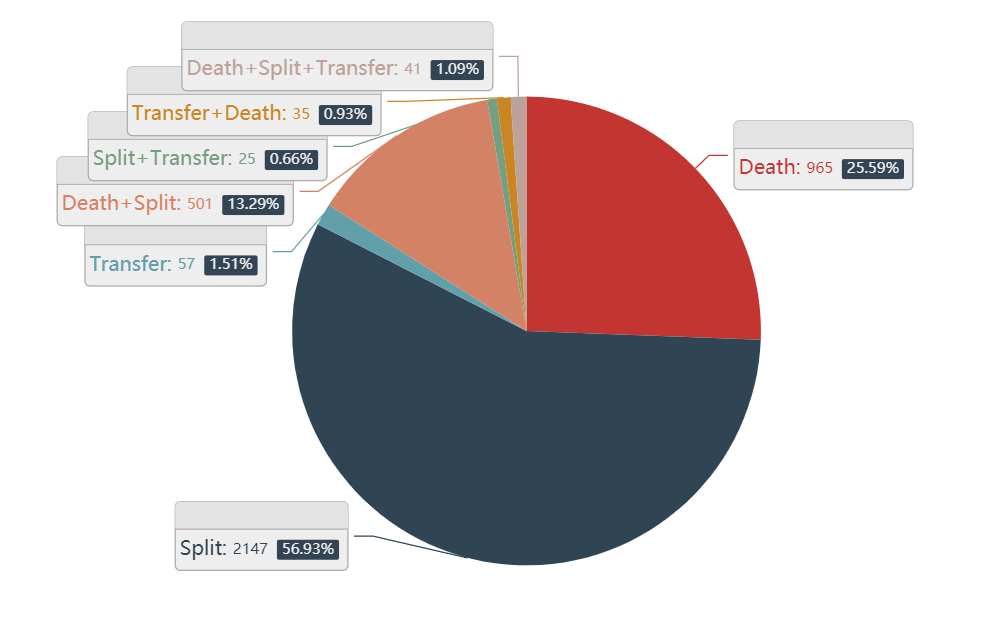}\label{fig:mashup_api_death_b}
}
\caption{Ratio of mashups that invoke different types of unavailable APIs.} \label{fig:mashup_api_death}
\end{figure}

Incorrect API composition information refers to mashups that are still available but are marked as invoking some unavailable APIs. We found $4,942$ mashups that invoked unavailable APIs, and $1,934$ that were still available. Figure~\ref{fig:mashup_api_death} shows mashups invoking the different types of unavailable APIs, where each sector is labeled with the number and percentage of mashups using a particular type of unavailable API. For example, in Figure~\ref{fig:mashup_api_death_b}, the number of available mashups invoking only \textit{split} APIs is $2,174$, which is $56.9\%$ of the number of available mashups that invoke unavailable APIs. The figure provides some insights that have been overlooked in the past studies:
\begin{enumerate}
\item APIs that are widely used and provide rich services tend to split and subsequently provide more refined services.
\item The importance of different APIs in mashups is different. Some APIs do not participate in a mashup or could be replaced by other APIs in the mashup. Thus, even if these APIs are abandoned, the mashup can still guarantee the availability of services.
\end{enumerate}

\subsection{Other Quality Issues}
This section discusses several other data quality issues in the original \textit{ProgrammableWeb} dataset. These issues are not relevant to the evolutionary analysis but are relevant to other tasks such as API tag prediction. This section also covers some quality issues that cannot be corrected without external data or manpower.It should be noted that these data quality issues do not affect the data correction methods in Section~\ref{sec:corrected}: 1) these issues are independent of the quality issues raised in Section~3.1, Section~3.2 and Section~3.3; 2) these issues are evolutionary-independent.

The descriptive texts of APIs and mashups contain considerable noise, and the information in the texts is sparse, which hinders tasks that require semantic support, such as service tag prediction \cite{gan2019multi} and service recommendation \cite{ma2020deep}. Bai et al. \cite{bai2017sr} also reported this problem and tried to use service representation-latent Dirichlet allocation (SR-LDA) to mine API/mashup topic words to alleviate the noise and sparsity problems. However, based on the results of their study, there is still much room for improvement.
The issue of the quality of the description text will not have an impact on the research questions or methods in this paper, as it is only private property of API/mashup that is independent of time and API/mashup status.

Toy mashups refer to the mashups that do not offer practical application value to users. Such mashups are usually the result of developers practicing or are created just for fun, such as \textit{Website-Grader}\footnote{https://www.programmableweb.com/mashup/website-gradercom} and \textit{Trump's Circle of Death}\footnote{https://www.programmableweb.com/mashup/trumps-circle-death}. Generally, these toy mashups should be treated as noisy data and filtered out. However, much manual annotation is required to accurately identify these toy mashups. Therefore, this issue is usually overlooked in research. Nevertheless, we note that toy mashup noise has a potential impact for algorithms operating on the entire dataset and that the extent of this influence depends on the application scenario. Fortunately, the toy mashups have no effect on most of the research questions in this paper because these research questions use only part of the data and automatically filter out the toy mashups. The toy mashups do have a small impact on RQ1 in Section~\ref{sec:RQ}, but they do not affect the conclusion drawn for the question.

A mashup is formed by combining various APIs in a certain sequence. Unfortunately, no relevant information exists regarding the combination sequence in the original \textit{ProgrammableWeb} dataset. Moreover, the original sequence is almost impossible to automatically recover based on information stored in the original \textit{ProgrammableWeb}. Therefore, we make a compromise and focus on the \textit{co-occurrences} of APIs in mashups.

\subsection{Impact of Data Quality Issues}
The implications of the abovementioned data quality issues for service computing tasks, except for the evolution analysis, are discussed in this section; implications for the evolution analysis are explained in detail after the answers to each of the research questions in Section~\ref{sec:RQ}.

Incorrect API/mashup availability status will significantly affect the reliability of the results generated by application scenarios, such as service recommendation, service selection, service combination, etc., that need to satisfy user requirements. Since they will recommend services that have been unavailable for a long time, this obviously does not meet the needs of realistic conditions.

Incorrect death time and availability status can be a major threat to the correctness of various time-aware methods. The original \textit{ProgrammableWeb} dataset is presented as an incremental dynamic network in which only new nodes (APIs, mashups) and edges (invoke relations) are added. However, the fact that the API/mashup fails at different times makes the data structure of an incremental dynamic network unrealistic, and the nodes and edges in this network can exit the network. This causes many operations that could be performed on an incremental dynamic network to be inoperable or costly in terms of time and space.

Incorrect API composition information, the lack of API combination sequence information, and toy mashups will affect the reliability of the results of tasks such as service composition, service discovery and service selection because these tasks rely on the statistical results of known API compositions
, and these data quality issues lead to unreliable and missing API composition profiles.

\section{Data Correction}\label{sec:corrected}
In this section, the data quality issues mentioned in Section~\ref{sec:problems} are addressed to better support the evolution analysis of the service ecosystem. Before expounding on the details of the method, we 
wish to clarify the following points:
\begin{itemize}
\item Our method is not perfect in the sense that the data correction results are not totally accurate; they are a probability estimate.
\item Because this paper focuses on service ecosystem evolution, our method aims at the whole service ecosystem; in other words, we can guarantee the rationality of the overall distribution of the data but cannot guarantee the accuracy of individual samples.
\item To balance the algorithm's complexity, human cost, and other factors, we ignored some unimportant 
factors and made several reasonable 
assumptions.
\end{itemize}

\subsection{Estimated Death Time}\label{sec:estimate}
\begin{figure}[!tbp]
\centering
\includegraphics[width=\linewidth]{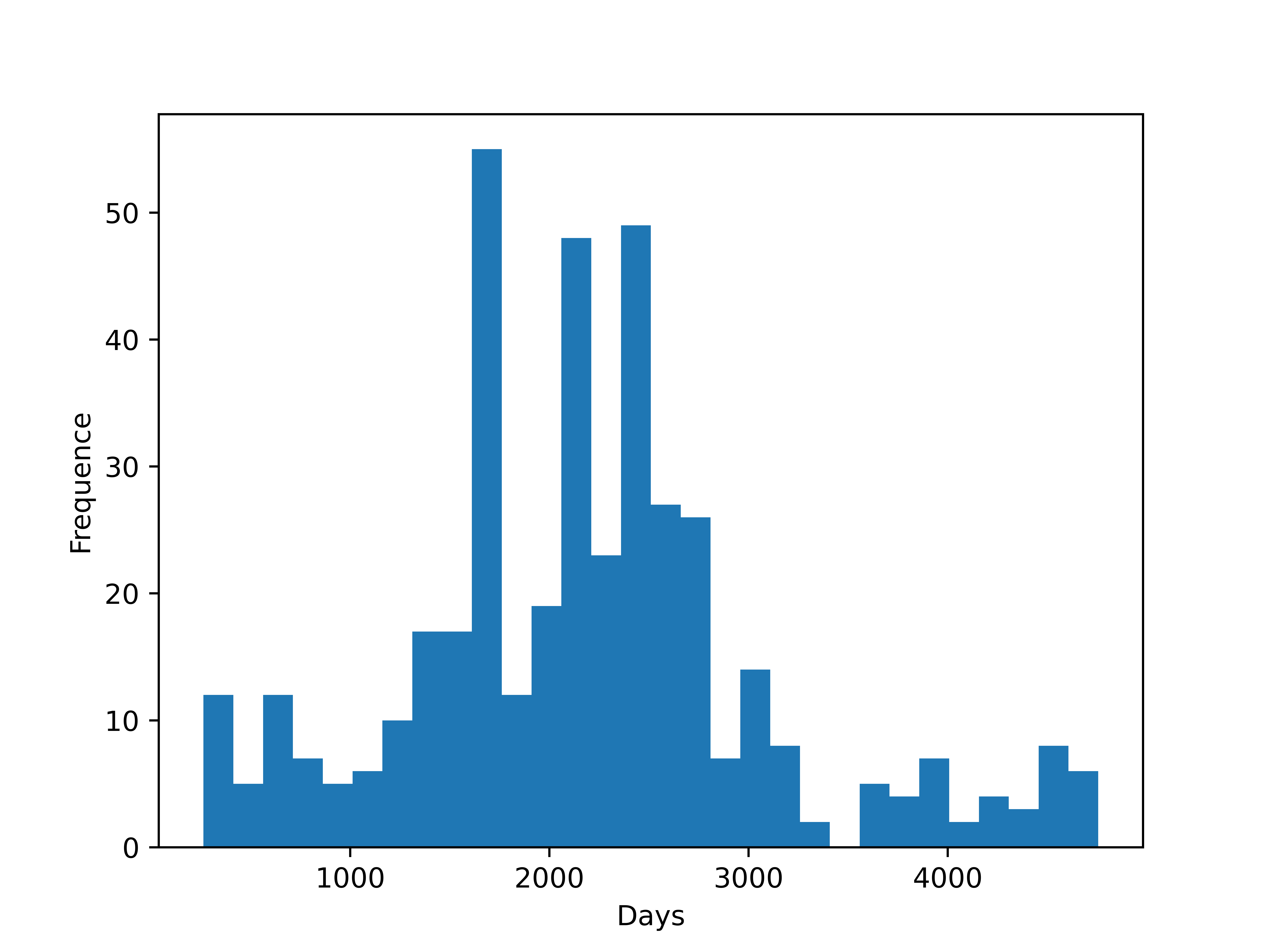}
\caption{The distribution of survival days for APIs placed in the deathpool from 2018 to 2020.} \label{fig:survival_days}
\end{figure}

It is not feasible to directly estimate the death times of APIs and mashups. Therefore, we calculated the death times by estimating survival days. The statistical results of the longevity of APIs placed in the deathpool from 2018 to 2020 are shown in Figure~\ref{fig:survival_days}.

Based on Figure~\ref{fig:survival_days}, we assume that the longevity of APIs and mashups $(x_1, x_2, \dots, x_n)$ follows a normal distribution $\mathcal{N} ( \mu ,\sigma^2 )$, which is one of the most common and applicable distributions in the real world. Then, the key problem is how to learn the approximate values of parameters $\mu$ and $\sigma^2$ based on the sample $(x_1, x_2, \dots, x_n)$. The standard approach for addressing this problem is to use the maximum likelihood method, which requires maximization of the \textit{log-likelihood} function:
\begin{equation}
\begin{aligned}
\ln \mathcal{L}(\mu, \sigma^2 ) &= \sum_{i=1}^n \ln f(x_i| \mu, \sigma^2) = -\frac{n}{2}\ln(2\pi) - \frac{n}{2}\ln\sigma^2 \\
&\quad- \frac{1}{2\sigma^2}\sum_{i=1}^n(x_i - \mu)^2
\end{aligned}
\end{equation}
where $f(x)$ is the probability density function (PDF). Taking the derivatives of $\mu$ and $\sigma^2$ and solving the resulting system of first-order conditions yields the maximum likelihood estimates:
\begin{equation}\label{equ:mu}
\hat{\mu} = \bar{x} \equiv \frac{1}{n}\sum_{i=1}^nx_i,
\end{equation}

\begin{equation}\label{equ:sigma}
\hat{\sigma}^2 = \frac{1}{n}\sum_{i=1}^n(x_i - \bar{x})^2.
\end{equation}

To ensure that the distribution estimated using the longevity of APIs placed in the deathpool from 2018 to 2020 is representative, we first randomly selected 200 unavailable APIs/mashups and manually checked their survival days \\$(x_1', x_2', \dots, x_n')$ through the Web snapshots provided by \textit{Wayback Machine}\footnote{http://web.archive.org} to represent the overall distribution. Then, we estimate the values of parameters $\hat{\mu}'$ and $\hat{\sigma'}^2$ based on the sample $(x_1', x_2', \dots, x_n')$ using Eq.\eqref{equ:mu} and Eq.\eqref{equ:sigma}. Finally, we use the Z-statistic to test whether the two distributions are the same:
\begin{equation}
Z = \frac{\hat{\mu} - \hat{\mu}'}{\sqrt{\hat{\sigma}^2+\hat{\sigma'}^2}}
\end{equation}
Normally, in qualitative terms:
\begin{itemize}
\item If the Z-statistic is less than 2, the two distributions are the same. 
\item If the Z-statistic is between 2.0 and 2.5, the two distributions are marginally different.
\item If the Z-statistic is between 2.5 and 3.0, the two distributions are significantly different.
\item If the Z-statistic is more then 3.0, the two distributions are highly significantly different.
\end{itemize}
The Z-statistic for these two distributions is $0.385$, which is much less than 2. Therefore, we can state that the distribution estimated using the longevity of APIs is the same as the distribution of the whole dataset, which means that the data in Figure 6 are representative.


For each obsolete API or mashup $x$, we generate a longevity value, $d_x$, from the normal distribution $\mathcal{N}(\hat{\mu}, \hat{\sigma}^2)$. The unavailable time $end_x$ can be calculated as follows:
\begin{equation}
end_x = 
\begin{cases}
start_x + d_x & \text{if } start_x + d_x <= \beta_x \\
random(start_x, \beta_x)& \text{otherwise},
\end{cases}
\end{equation}
where $start_x$ is the creation time of $x$; $\beta_x$ denotes the earliest time that $x$ is confirmed to be unavailable. 
In this paper, we take the value as the date of the first network request test (September 10, 2020). We also introduce a random distribution $random(\cdot)$ to prevent $end_x$ from being later than $\beta_x$. In particular, with the \textit{transferred/split} APIs, we do not have to use such a method to estimate the unavailable time. Suppose an API $x$ has \textit{split/transferred} into a new API set $\{x_1, x_2, \dots, x_m\}$. 
The unavailable time of $x$ is:
\begin{equation}
end_x = \max(start_{x_1}, start_{x_2}, \dots, start_{x_m}).
\end{equation}

\subsection{Correct API Composition for Mashups}

We use $M_t = \{a_1, a_2, \dots, a_k\}$ to represent an available mashup at time $t$, where $a$ denotes an API included in the mashup. Suppose at time $t+1$ the mashup $M_{t+1}$ is still available, but API $a_i$ is no longer available. The representation of $M_{t+1}$ depends on the pattern of $a_i$ being unavailable, as shown in Equation~\eqref{equ:api_composition}:
\begin{equation}
M_{t+1} = 
\begin{cases}
M_t - \{a_i\} & \text{if \textbf{death}} \\
M_t - \{a_i\} + \{a'_i\} & \text{if \textbf{transfer}}, a_i \rightarrow a'_i \\
M_t - \{a_i\} + \{a_{i,1}, \dots, a_{i, n}\} & \begin{aligned}
&\text{if \textbf{split}},\\
&a_i \rightarrow \{a_{i,1}, \dots, a_{i, n}\}
\end{aligned}
\end{cases}
\label{equ:api_composition}
\end{equation}

The handling of \textit{death} and \textit{transfer} cases is easy to understand. For \textit{split} APIs, we made a reasonable 
assumption due to imperfections in the information. We assumed that all the APIs $a_{i,j}$ generated by splitting will be used by the mashup to ensure complete functionality despite knowing that only one subset 
would be used in reality.
For example, mashup \textit{Mosoto}\footnote{https://www.programmableweb.com/mashup/mosoto} was composed of \textit{Box}\footnote{https://www.programmableweb.com/api/box} and \textit{Facebook}\footnote{https://www.programmableweb.com/api/facebook} services when it was first created. Thus, it can be denoted as \{\textit{Box}, \textit{Facebook}\}.
Then, something happened on \textit{Facebook} that caused it to split into \textit{Facebook Ads}\footnote{https://www.programmableweb.com/api/facebook-ads}, \textit{Facebook Atlas}\footnote{https://www.programmableweb.com/api/facebook-atlas}, \textit{Facebook Graph}\footnote{https://www.programmableweb.com/api/facebook-graph} and \textit{Facebook Marketing}\footnote{https://www.programmableweb.com/api/facebook-marketing}. Subsequently, the representation of \textit{Mosoto} is \{\textit{Box}, \textit{Facebook Ads}, \textit{Facebook Atlas}, \textit{Facebook Graph}, and \textit{Facebook Marketing}\}.
Next, when \textit{Facebook Ads} and \textit{Facebook Atlas} are deprecated, the \textit{Mosoto} representation changes again to \{\textit{Box}, \textit{Facebook Graph}, and \textit{Facebook Marketing}\}.

\section{Service Ecosystem Dynamic Network Model}\label{sec:model}
The service ecosystem is a continuously evolving complex network system consisting of service entities and interactions between them. This system can be naturally modeled as three different dynamic networks in which the structure changes over time depending on different scenarios:
i) the Mashup-API network (M-A), ii) the API-API network (A-A), and iii) the category-category network (C-C).

\begin{definition}
The Mashup-API network (M-A) is a dynamic bipartite graph $G_{MA} = \{A, M, E_{MA}\}$, where $A$ refers to the APIs, $M$ refers to the mashups, $E_{MA}=\{(u, v, start, end)| u \newline \in M, v \in A\}$ denotes the invoking relations between mashups and APIs, and $start$ and $end$ represent the duration of each relation. An API/mashup $x \in A \cup M$ can be denoted as $x = \{start, end, c\}$, where $c$ represents the primary category to which $x$ belongs. $G^t_{MA} = \{A^t, M^t, E^t_{MA}\}$ is used to represent a snapshot corresponding to the M-A at time $t$.
\end{definition}

\begin{definition}
The API-API network (A-A) is a homogeneous network serial snapshot, denoted as $G_{AA} = \{G^1_{AA}, G^2_{AA}, \dots, \\ G^n_{AA}\}$. 
$G^t_{AA} = \{A^t, E^t_{AA}\}$ is the A-A snapshot at time $t$ generated by a Mashup-API network snapshot $G^t_{MA}$. $E^t_{AA} = \{(u, v, w) | u, v \in A^t\}$ represents the API concurrence relations, and $(u, v, w)$ indicates that API $u$ and API $v$ are invoked together in an available mashup $w$ at time $t$.
\end{definition}

\begin{definition}
The category-category network (C-C) is a hypergraph serial snapshot, denoted as $G_{CC} = \{G^1_{CC}, G^2_{CC}, \dots\\, G^n_{CC}\}$. 
$G^t_{CC} = \{C^t, E^t_{CC}\}$ is the C-C snapshot at time $t$ generated by an API-API network snapshot $G^t_{AA}$, where $C$ is a set of categories to which the APIs belong. $E^t_{CC} = \{(u, v, w) | u, v \in C\}$ indicates the number of times that an API of category $u$ and an API of category $v$ are invoked together in an available mashup $w$ at time $t$.
\end{definition}

\begin{figure*}
\centering
\includegraphics[width=0.8\textwidth]{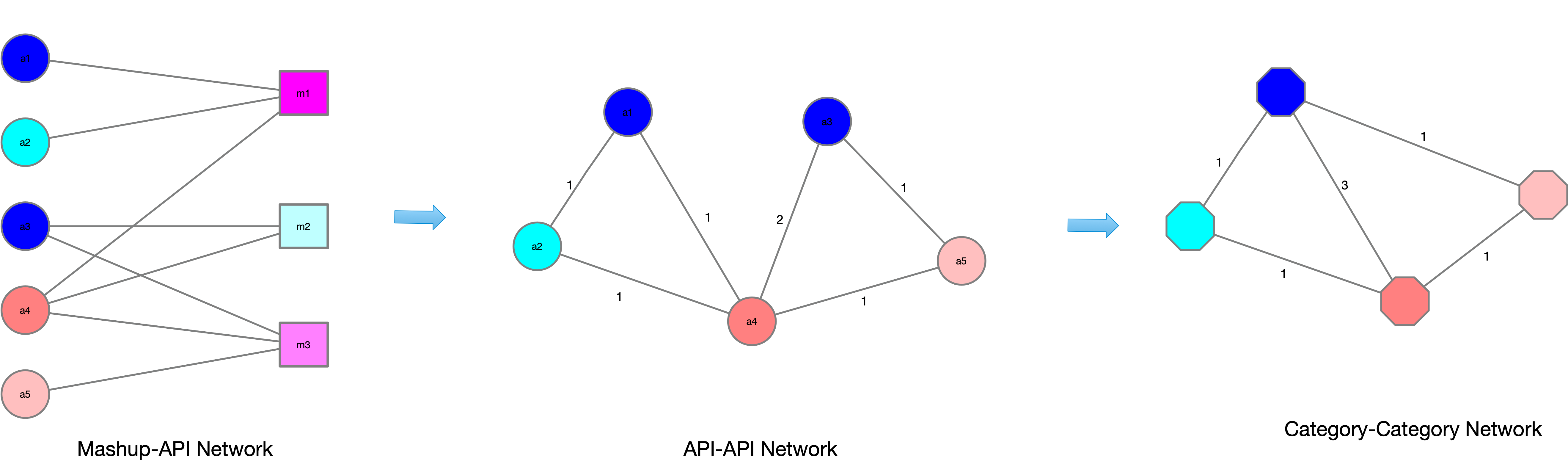}
\caption{Conversion between the three dynamic network types.} \label{fig:model_transition}
\end{figure*}

Figure~\ref{fig:model_transition} shows the conversion process between the three dynamic network types. The square in the figure represents a mashup, the circle represents an API, and the color of the shape denotes the primary category. The link in the figure denotes the invoking/concurrence relations. These three different dynamic networks reflect the evolution of the \textit{ProgrammableWeb} service ecosystem from different levels, which will be discussed in detail in the next section.

\section{Service Ecosystem Evolution Analysis}
\label{sec:RQ}
\begin{figure}
\centering
\includegraphics[width=0.9\linewidth]{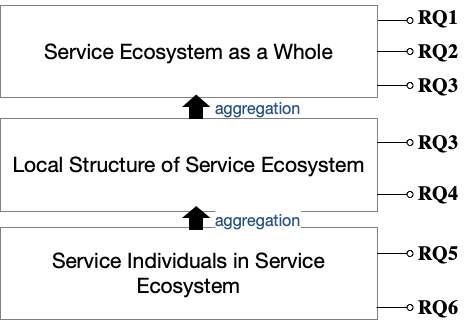}
\caption{The link between six research questions.}
\label{fig:rq_relations}
\end{figure}
This section analyzes the service ecosystem evolution from the perspective of complex networks. The subsequent analysis in this section is organized by answering six research questions and identifying research challenges and opportunities based on the analysis. The title of each research question consists of two parts. The first part concerns which dynamic network model is selected to support the research question, and the second part explains the research question. 
As shown in Figure~\ref{fig:rq_relations}, we have constructed 6 research questions at three levels: the service ecosystem as a whole (\textbf{RQ1, RQ2, RQ3}), the local structure of the service ecosystem (\textbf{RQ3, RQ4} and service individuals in the service ecosystem (\textbf{RQ5, RQ6}). In particular, in \textbf{RQ1, RQ2, RQ4 and RQ5}, we found different conclusions from existing studies. Furthermore, in \textbf{RQ3} and \textbf{RQ6} we found that the current \textit{ProgrammableWeb} service ecosystem cannot satisfy the complex service market requirement.

It is worth mentioning 
that although the method 
discussed 
in Section~\ref{sec:corrected} cannot guarantee that an API/mashup individual is completely correct, it does guarantee the accuracy of the {\em overall} distribution of the \textit{ProgrammableWeb} service ecosystem. 
We repeated the experiments several times independently and found that the results have no influence on the conclusions of the evolutionary analysis in this paper.

\vspace{2mm}
\noindent\textbf{RQ1: [M-A] What is the health status of the \textit{ProgrammableWeb} service ecosystem?}

\begin{figure}[!tbp]
\centering
\subfigure[APIs]{
\includegraphics[width=0.75\linewidth]{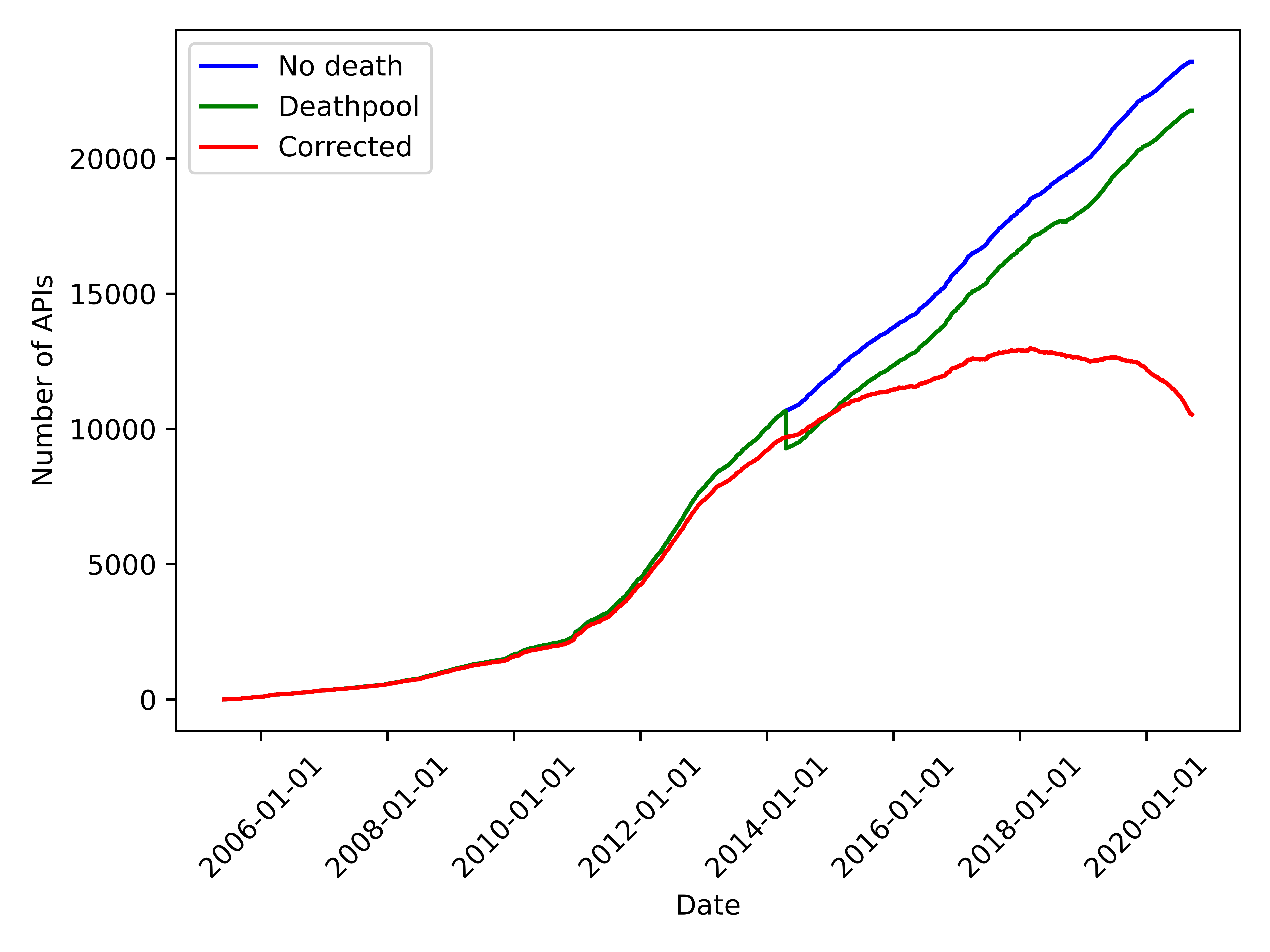}\label{fig:mashup_api_number_a}
}
\subfigure[mashups]{
\includegraphics[width=0.75\linewidth]{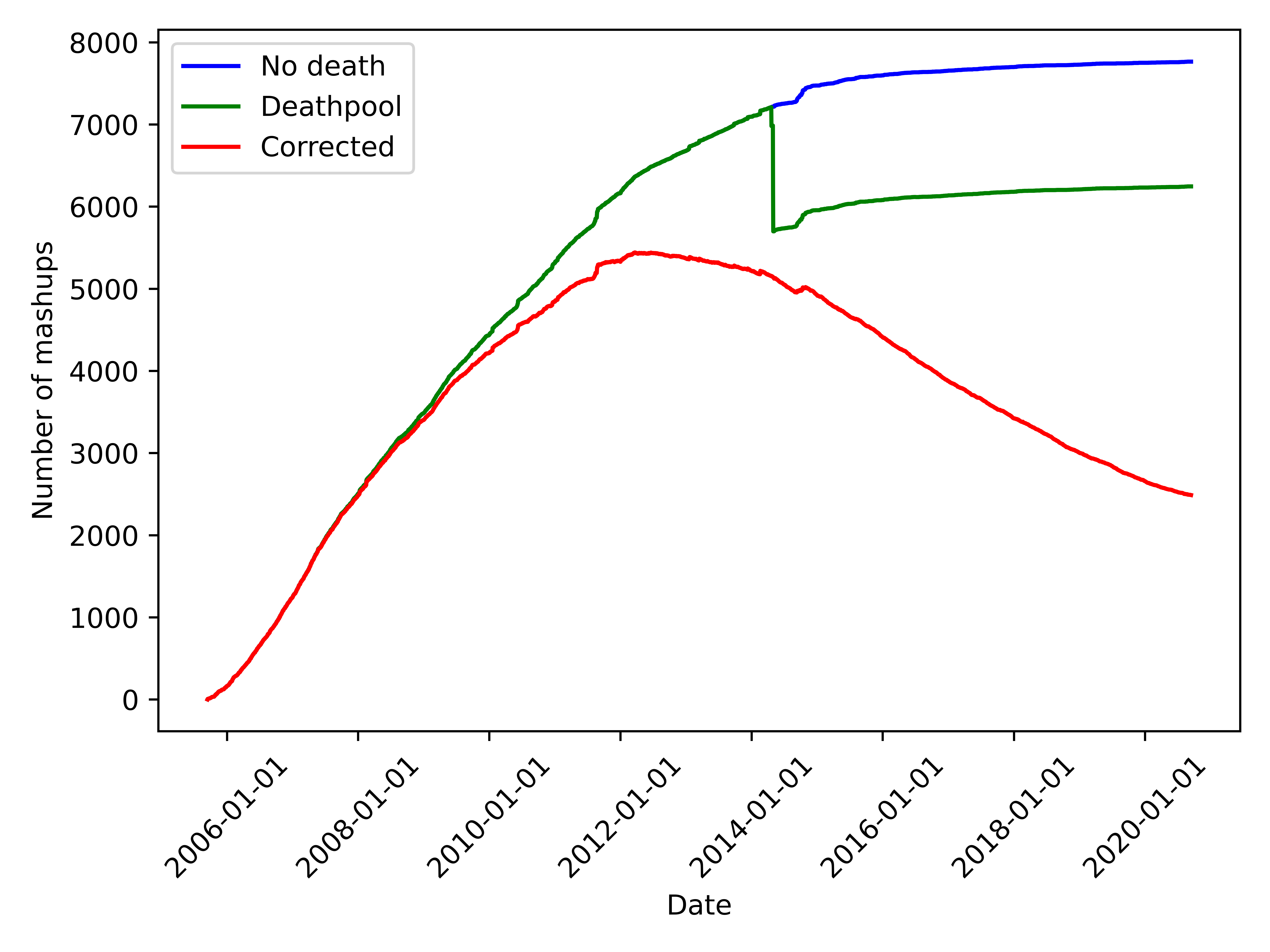}\label{fig:mashup_api_number_b}
}
\caption{The numbers of available APIs and mashups change over time (daily).} \label{fig:mashup_api_number}
\end{figure}

The number of available APIs and mashups is the most intuitive indicator of the health and prosperity of a service ecosystem. Figure~\ref{fig:mashup_api_number} shows how the number of available APIs and mashups changes over time. The blue line 
(i.e., No death) denotes that APIs and mashups are always available; the green line (i.e., Deathpool) denotes using the deathpool information provided by the original \textit{ProgrammableWeb} to determine whether an 
API or mashup 
is available; and the red line (i.e., Corrected) denotes the results of our corrected \textit{ProgrammableWeb} dataset.

To the best of our knowledge, all the existing studies are based on the first two scenarios (No death or Deathpool), and consequently, they provide ``optimistic'' estimates regarding the health and prosperity of the \textit{ProgrammableWeb} service ecosystem \cite{tian2017exploratory}. 
However, our corrected data show that the reality might be much worse.

Based on the change curve of the number of available APIs and mashups, there is no evidence to support continuous ``explosive'' growth, which has often been mentioned in the existing studies. The results show that after 2014, the \textit{ProgrammableWeb} service ecosystem actually began to decline. Although the number of new APIs remains stable year over year, the rate of API obsolescence has continued to accelerate, especially after 2018. While the addition of new mashups has stagnated since 2014, the rate of obsolescence has remained steady. We argue that the reason for this phenomenon is that the popularity of the mobile Internet has led to an increasing shift of Web mashups to mobile apps~\cite{WangWXS17}, which is also 
supported by online statistics\footnote{https://www.statista.com/statistics/266210/number-of-available-applications-in-the-Google-play-store/}.

In summary, the changes in the number of mashups and APIs in the \textit{ProgrammableWeb} service ecosystem indicate that \textbf{the ecosystem is in poor health and could become extinct unless necessary steps are taken to remedy the situation}.

\vspace{2mm}
\noindent\textbf{RQ2: [A-A] Does the degree distribution of the service ecosystem network comply with the power law?}

In almost all studies on service ecosystem network evolution, {\em degree distribution} is a core element. Many studies \cite{10.1007/978-3-030-03596-9_44, adeleye2019fitness, lyu2014three, zhou2019service} have concluded that the degree distribution of the A-A network complies with the power law, and a generative model of the A-A network is built to support downstream tasks. However, based on the theoretical analysis and our experimental results, we believe that this conclusion may be misleading due to insufficient statistics and inappropriate evaluation indicators.

\begin{figure*}
\centering
\includegraphics[width=0.95\textwidth]{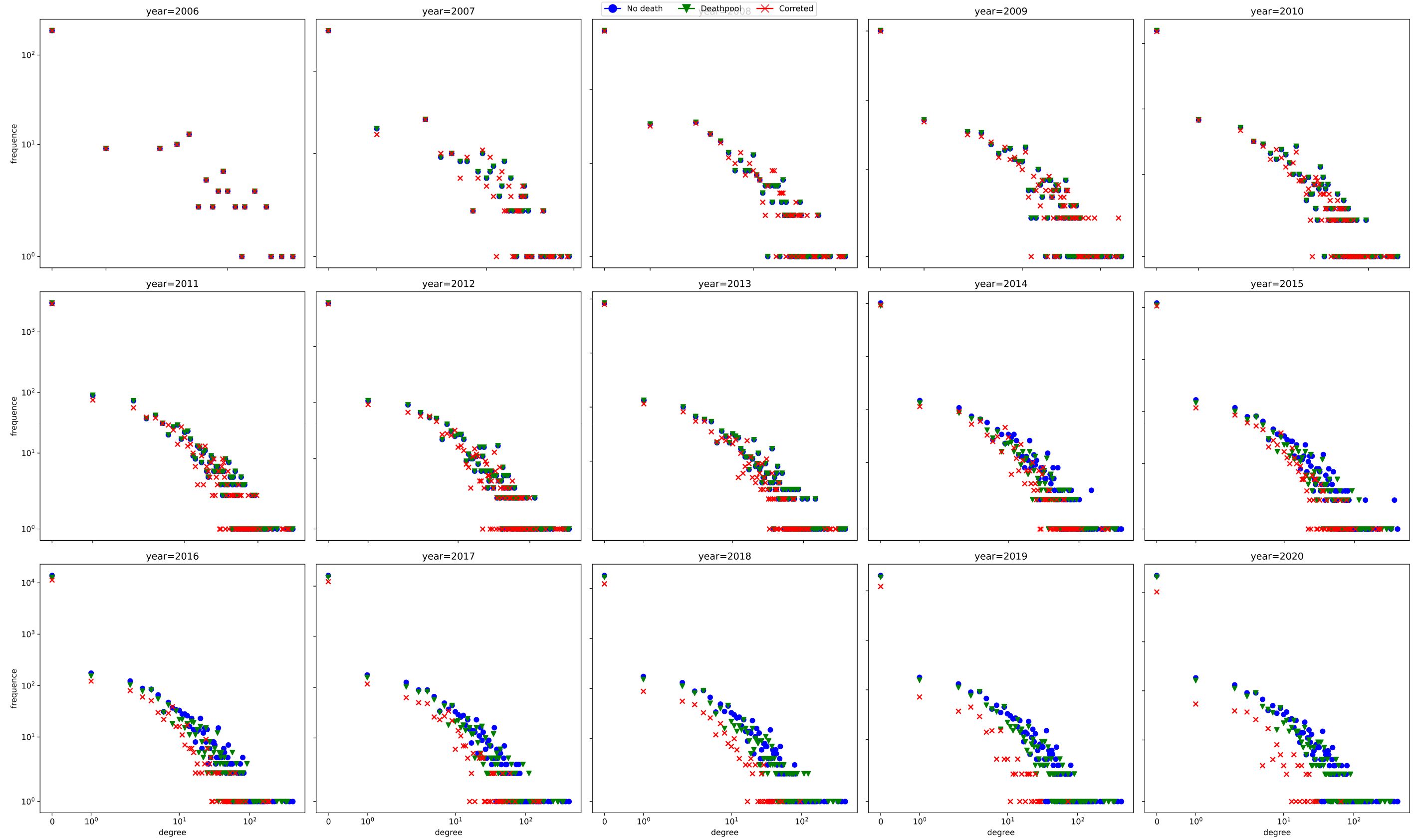}
\caption{The degree distribution changes over time (yearly).}\label{fig:degree_distribution}
\end{figure*}

\begin{figure}
\centering
\includegraphics[width=0.8\linewidth]{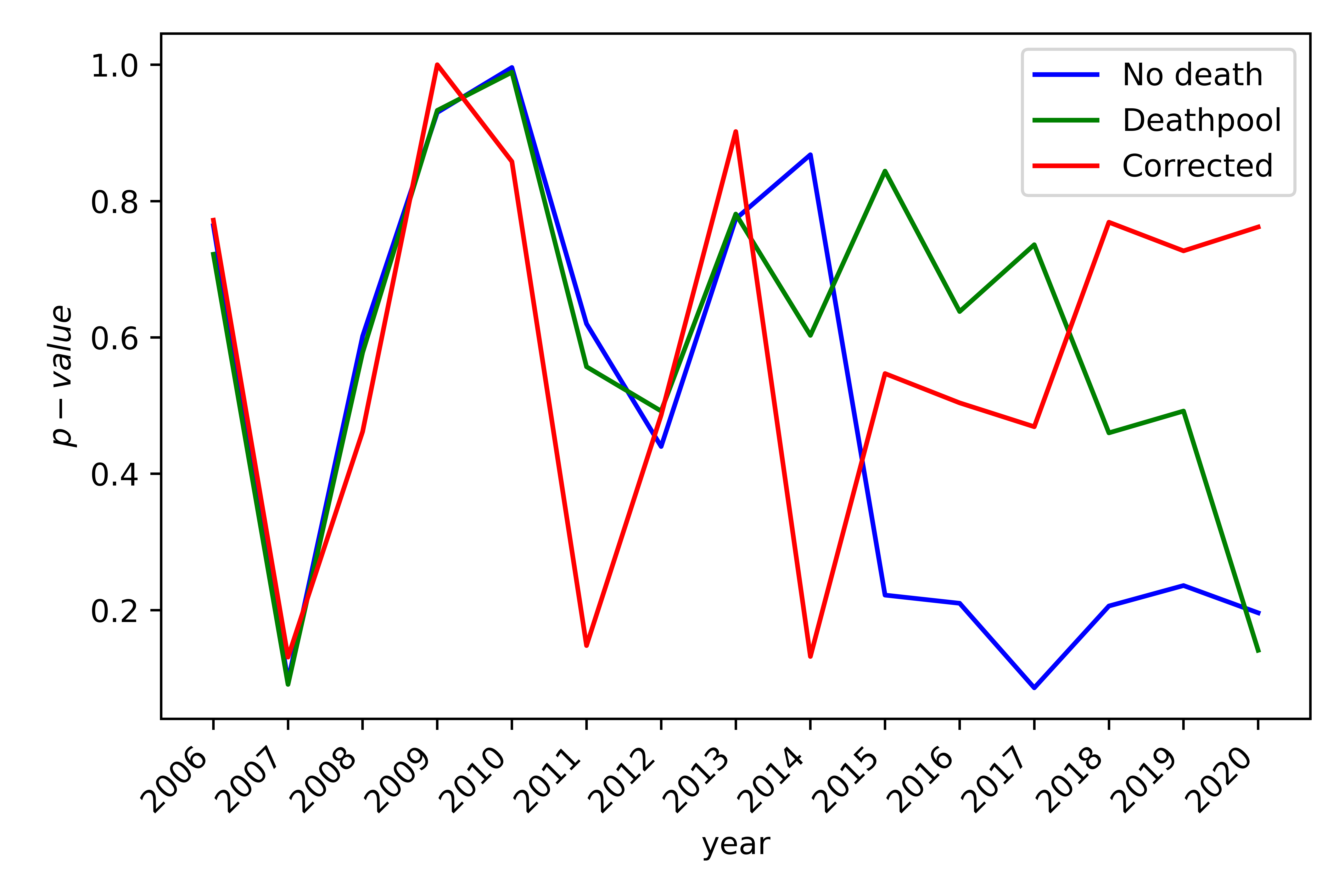}
\caption{The $p$-$value$ for the power law changes over time (yearly).}\label{fig:power_law}
\end{figure}

Figure~\ref{fig:degree_distribution} presents the results of the degree distribution changes over time. 
The figure shows that the degree distribution is quite consistent with the power 
law\footnote{If consistent, the overall trend should be close to a straight line, not a curve.}, especially on the corrected \textit{ProgrammableWeb} dataset.
The existing studies usually adopt the $p$-$value$ as evidence that the degree distribution of the \textit{ProgrammableWeb} service ecosystem network conforms to the power law. We first followed \cite{10.1007/978-3-030-03596-9_44} and conducted a \textit{goodness-of-fit} test based on the Kolmogorov-Smirnov (KS) distance to calculate a $p$-$value$ to quantitatively measure the plausibility of the power law. The results are shown in Figure~\ref{fig:power_law}.

We can see from Figure~\ref{fig:power_law} that the $p$-$value$ fluctuates substantially. Thus, we argue that the results do not support that the power law is a plausible hypothesis for the data:

\begin{itemize}
\item A $p$-$value$ depends upon both the magnitude of association and the precision of the estimate (the sample size), while a smaller sample size tends to result in a larger $p$-$value$. $p$-$values$ are typically used to reach a conclusion of ``significant'' or ``not significant'' based on whether the $p$-$value$ is larger than a threshold. Consequently, $p$-$values$ are more similar to qualitative measurements than quantitative measurements. For a detailed discussion we refer readers to \cite{clauset2009power}.
\item $P$-$values$ are computed based on the assumption that the null hypothesis is true. 
A $p$-$value$ represents the probability that the data deviates from the null hypothesis by a certain amount.
Consequently, a p-value measures the compatibility of the data with the null hypothesis---not the probability that the null hypothesis is correct. Therefore, it is unreasonable to compare $p$-$values$ associated with different null hypotheses as in \cite{10.1007/978-3-030-03596-9_44}.
\end{itemize}

\textbf{To summarize, we argue that the current statistics and measurement methods are insufficient, inappropriate, or have difficulty supporting the hypothesis that the degree distribution complies with the power law.} Downstream tasks such as service recommendation and service discovery based on the existing findings may risk incorrect hypotheses and degrade the effectiveness of the method.

\vspace{2mm}
\noindent\textbf{RQ3: [C-C] What are the changes in diversity and popularity of different categories in the service ecosystem?}

\begin{figure*}[!tb]
\centering
\includegraphics[width=\textwidth]{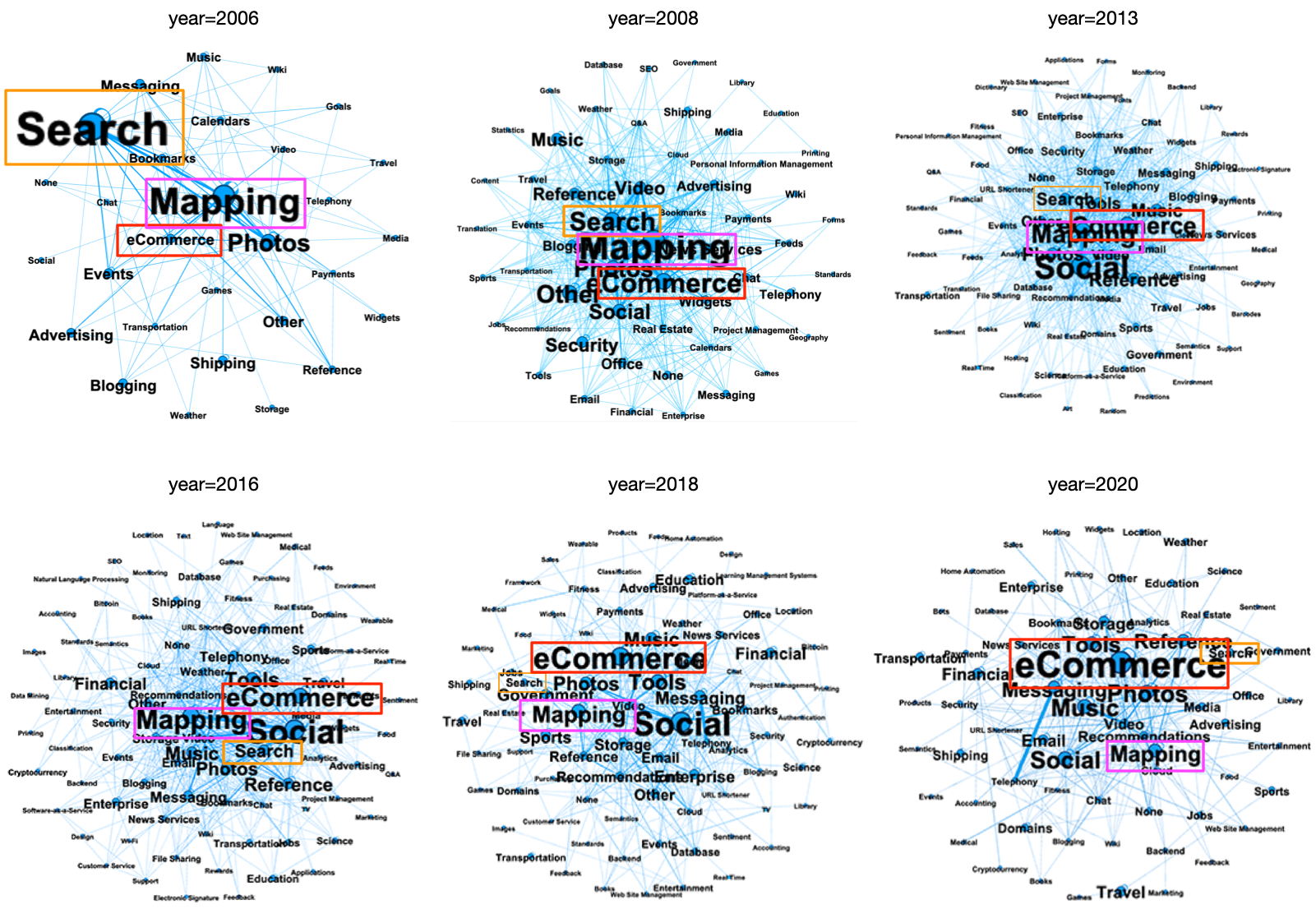}
\caption{Visualization of the evolution of the category-category network (yearly).}\label{fig:topic_evolve}
\end{figure*}
Diversity reflects the vitality of an ecosystem and is also positively related to the richness of the services that a service ecosystem can provide \cite{huang2013impact, kochan2003effects}. Understanding the changing trends of popular service categories in a service ecosystem can provide developers with suggestions when providing new services. The C-C network naturally reflects the diversity 
and the popularity of various categories in the service ecosystem. 
The number of nodes indicates the diversity of service categories, and the locations (generated by visual layout) and sizes of nodes (number of aggregated APIs) can be used to indicate popularity.

\begin{figure}[!h]
\centering
\includegraphics[width=0.8\linewidth]{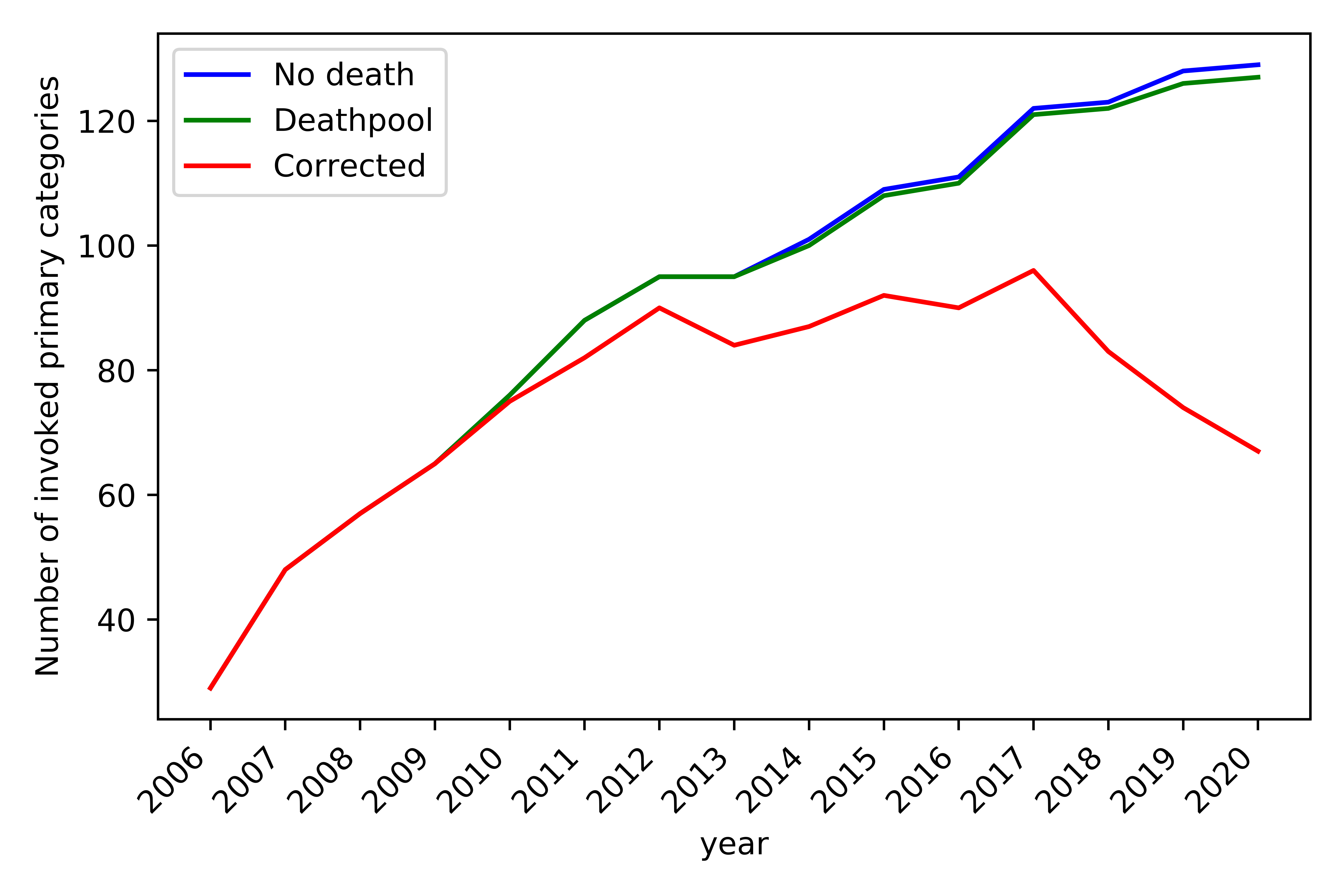}
\caption{The number of invoked primary categories (diversity) changes over time (yearly).}\label{fig:diversity}
\end{figure}

Figure~\ref{fig:diversity} depicts the diversity of the \textit{ProgrammableWeb} service ecosystem changes over time by counting the number of nodes in the C-C network, which is generated from the corresponding A-A network after removing isolated APIs. \textbf{The red curve shows that the diversity of the \textit{ProgrammableWeb} service ecosystem first increased, then entered a plateau, and began to decline rapidly after 2017.}

The evolution of the C-C network is visualized using \textit{Gephi}\footnote{An open graph viz platform: https://gephi.org} and shown in Figure~\ref{fig:topic_evolve}, which also indicates the changes in the popularity of different categories. In particular, \textit{Mapping} has been a popular category from the beginning. Although the node size has undergone a relative decrease, it still captures the center place, which means that this category is often used with other categories (we all use such services everywhere). For example, when searching for a restaurant on a map, users also obtain additional information, such as customers' reviews and discounts.

\textit{eCommerce} also appeared from the beginning, but it did not undergo rapid development until recent years. However, its popularity has remained consistent. In recent years, eCommerce nodes have become one of the largest node clusters in the graph; that is, eCommerce is included as part of other categories' services more than ever before. Therefore, when creating a new mashup to provide a new service, the mashup may achieve a higher probability of recommendation if it includes eCommerce services.

Unlike the categories mentioned above, the \textit{search} node was initially very large and then gradually decreased. It also moved to the edge of the graph. This implies that using complete search services is not as necessary as it was before. Instead, services may implement their own searches rather than relying on search services provided by others.

Additionally, Figure~\ref{fig:topic_evolve} also reflects the slow update and incomplete inclusion of APIs and mashups in the current \textit{ProgrammableWeb} dataset. For example, the APIs of speech recognition, natural language processing (NLP), image processing and other cognitive services, which have been promoted by large companies such as \textit{Microsoft}, \textit{Google}, \textit{Amazon}, and \textit{Baidu} in recent years, have not been included.

\begin{figure}
\centering
\includegraphics[width=0.8\linewidth]{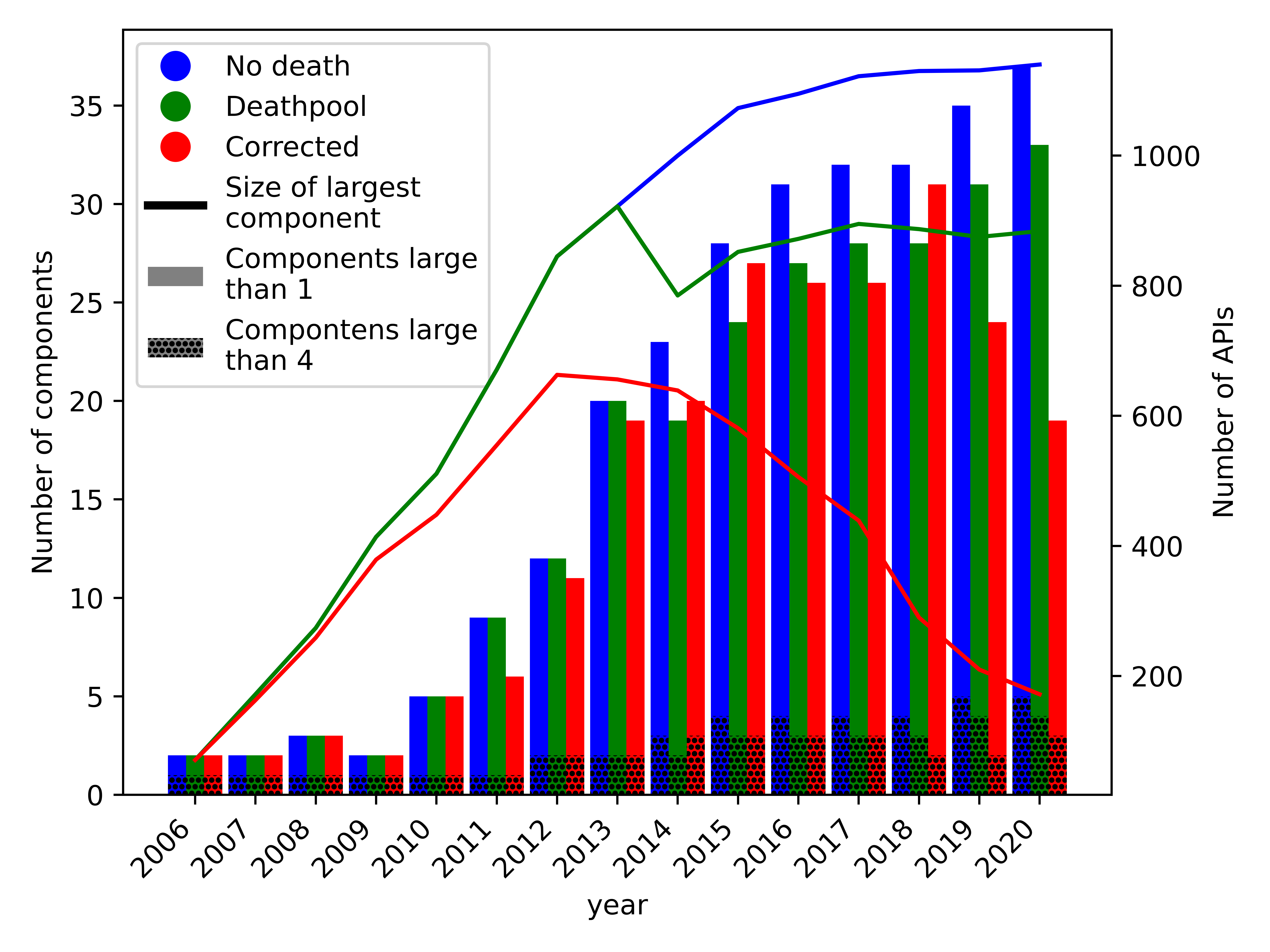}
\caption{The number of connected components and the size of the largest connected component (yearly).}\label{fig:components}
\end{figure}

\vspace{2mm}
\noindent\textbf{RQ4: [A-A] What are the changes in the number and size of connected components over time?}

The connected components form a subgraph in which any two vertices are connected to each other by paths that are connected to no additional vertices in the subgraph. Connected components are an important feature of the A-A network. On the one hand, connected components can reflect the development of the service ecosystem via their number and size. Intuitively,
we know that if the entire A-A network is a large connected component, more combinations are available, making it more likely that rich applications (mashups) will be created. 
On the other hand, many service recommendation and service discovery algorithms \cite{zhong2014time, gao2016joint, 10.1007/978-3-030-03596-9_44} are essentially performed on connected components that reach a certain scale.

Figure~\ref{fig:components} shows how the number of connected components and the size of the largest connected components has changed over time. \textbf{The total number of connected components first grew, then stabilized, and then began to decline sharply after 2018, but the number of components greater than 4 remained stable. The size of the largest connected component has continued to shrink since 2013; by 2020, it was below $200$}. The figure also shows that many isolated connected components exist in the A-A network and that most of the connected components are very small (less than $5$). From a data science perspective, most of the data (isolated small components) provide poor information when performing tasks such as data-driven service recommendation, service discovery, and service composition. The size of the largest connected component is also quite small, which introduces the small data problem and greatly limits the ability to apply advanced machine learning methods \cite{zhang2019graph, zhang2020deep, goyal2018graph, wang2018graphgan} to service computing.

\begin{table*}[!tb]
\caption{Top 5 most frequently co-occurring API pairs and 5 low-frequency but high-survival-rate API pairs.}\label{tab:api_pairs}
\begin{tabular}{|l|l|r|r|r|r|}
\hline
\bf API1 & \bf API2 & \bf Active Use & \bf Total Use & \bf Survival Rate & \bf Avg Days \\ \hline
/api/twitter & /api/google-maps & 63 & 185 & 0.34 & 1,405.09 \\ \hline
/api/youtube & /api/flickr & 36 & 184 & 0.2 & 1,589.16 \\ \hline
/api/twitter & /api/facebook & 64 & 179 & 0.36 & 1,414.26 \\ \hline
/api/google-maps & /api/flickr & 47 & 169 & 0.28 & 1,608.62 \\ \hline
/api/google-maps & /api/youtube & 48 & 157 & 0.31 & 1,596.5 \\ \hline
/api/google-visualization & /api/google-maps & 8 & 10 & 0.8 & 769.0 \\ \hline
/api/facebook & /api/instagram-graph & 9 & 13 & 0.69 & 1,094.75 \\ \hline
/api/instagram-graph & /api/twitter & 12 & 18 & 0.67 & 1,212.17 \\ \hline
/api/google-maps & /api/google-geocoding & 21 & 33 & 0.64 & 1,117.75 \\ \hline
/api/google-maps & /api/zillow & 7 & 11 & 0.64 & 1,070.25 \\ \hline
\end{tabular}
\end{table*}

\begin{figure}
\centering
\includegraphics[width=0.75\linewidth]{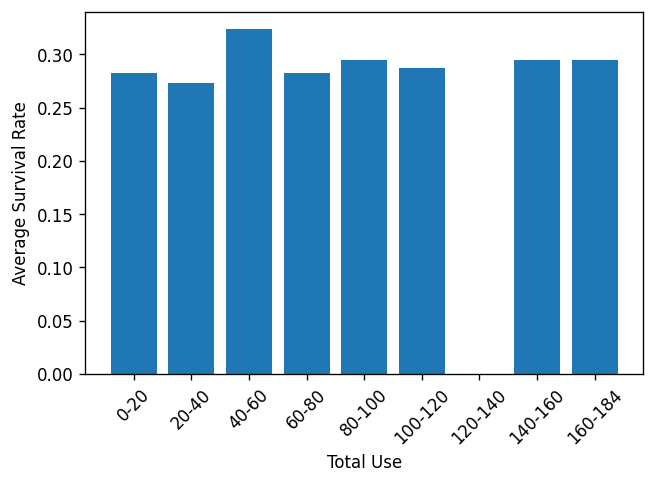}
\caption{Survival rate distribution of mashups that use co-occurring APIs.}
\label{fig:use_rate_avg}
\end{figure}

\vspace{2mm}
\noindent\textbf{RQ5: [A-A] Are mashups that use frequently co-occurring APIs more likely to survive?}

The existing service recommendation methods \cite{ma2020deep, gao2017novel} essentially involve statistical analysis and processing of the frequency of co-occurring APIs. The more frequently an API pair appears, the more likely it is to be recommended to developers. This naturally raises the question of whether frequently co-occurring APIs truly represent a good combination. More directly, is the survival rate of mashups that use frequently co-occurring APIs higher?

Figure~\ref{fig:use_rate_avg} demonstrates the changes of mashups' average survival rates over the co-occurring frequencies of APIs. The survival rate reaches the highest level when the frequency is between 40 and 60 and then decreases, which clearly indicates that using high-frequency pairs does not mean a tendency to survive.

Table~\ref{tab:api_pairs} lists some co-occurring APIs as examples: the first five are the most frequently co-occurring API pairs, and the last five are low-frequency 
(e.g., less than or approximately 30 times) but high-survival-rate API pairs. 
Clearly, the mashups that invoke the frequently co-occurring API pairs do not have a higher survival rate.
The survival rates of the five API pairs with the highest usage frequency are all below $40\%$, and those that use (\textit{YouTube} and \textit{Flickr}) are as low as $20\%$. However, the survival rate of some API combinations used at low frequencies 
exceeds $60\%$, and the highest reach $80\%$. Another interesting 
observation is that among the high-survival API pairs, in most cases, the two APIs are from the same developer, or one of the API categories is \textit{Social} (e.g., \textit{Twitter}, \textit{Facebook}, or \textit{Instagram Graph}).

In conclusion, \textbf{mashups that use frequently co-occurring API combinations do not mean that they are more likely to survive.} This phenomenon has brought new challenges to service recommendation and service composition, such as how to consider factors such as survival rate and developer relationships in recommendation or composition.

\vspace{2mm}
\noindent\textbf{RQ6: [M-A] Are new mashup sizes becoming larger over time?}

With the rapid development of the Internet, user requirements have become increasingly complex. It has become a trend for service providers to connect 
more 
external applications to their own platforms to satisfy these increasingly complex user requirements \cite{wu2015modern,xu2015big,weiss2010modeling}. For example, \textit{WeChat} and \textit{Alipay} achieved this by \textit{Mini-program}; \textit{Microsoft} released \textit{Office}\footnote{https://play.google.com/store/apps/details?id=com.microsoft.office.\\officehubrow} to enable users to access \textit{Word}, \textit{PowerPoint} and \textit{Excel} in one app. Therefore, a large size mashup tends to have more complete functionalities.

We wanted to know whether the abovementioned evolutionary trend is also occurring in Web application-based mashups, that is, whether mashups have changed to meet the increasingly complex requirements of users.

\begin{figure}[!btp]
\vspace{-5mm}
\centering
\subfigure[The complexity of new mashups changes over time. ]{
\includegraphics[width=0.8\linewidth]{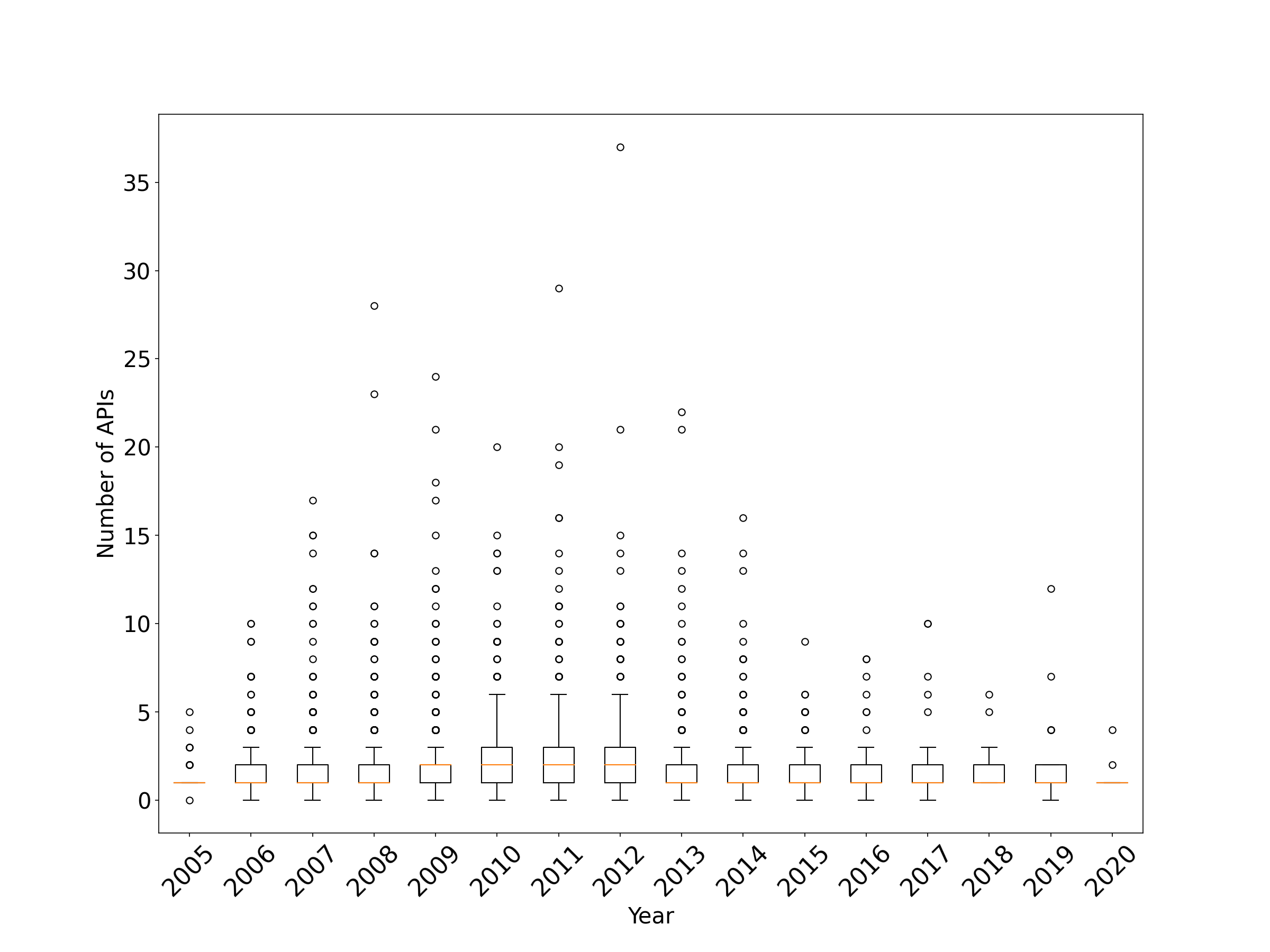}\label{fig:mashups_complexity}
}
\subfigure[The complexity of all mashups changes over time.] {
\includegraphics[width=0.8\linewidth]{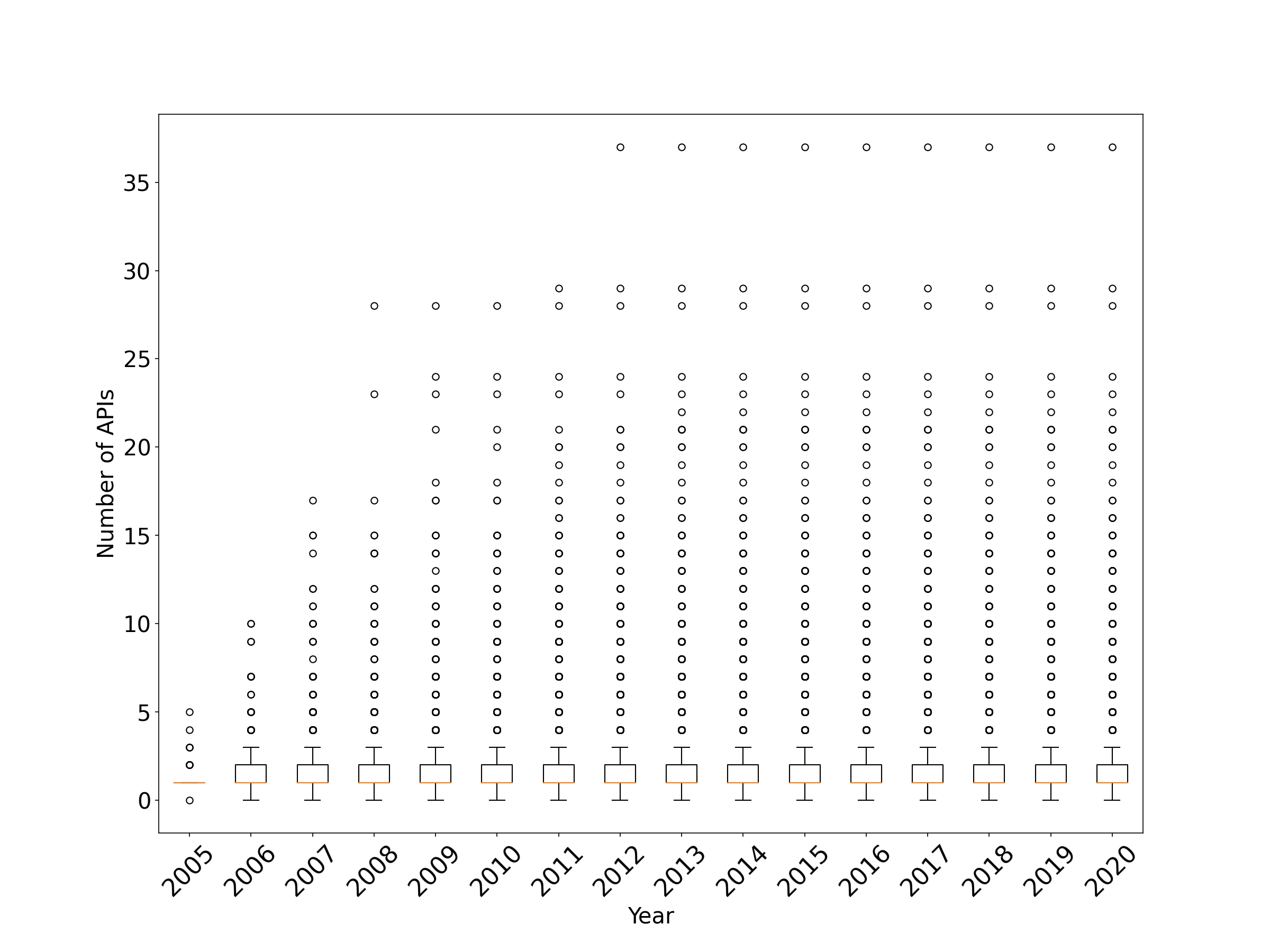}\label{fig:mashups_complexity_overall}
}
\caption{The complexity of mashups changes over time (yearly).}
\end{figure}

A mashup's size can be denoted by the number of APIs it invokes. Figure~\ref{fig:components} shows a box chart of the number of APIs invoked by newly submitted mashups each year. From the figure, we can see that most mashups provide only a simple service and the number of APIs they invoke is less than 3. The overall size of newly submitted mashups changed very little between 2006 and 2020: the average numbers of APIs they invoke remains steady between 1.2 and 2.5. 
The boxplot in Figure~\ref{fig:mashups_complexity_overall} shows the changes in size among all mashups. Apparently, no large change has occurred since 2012 in any part of this boxplot. Another interesting phenomenon is that the size of new mashups gradually increased each year prior to 2012. The most complex mashup added in 2012 was composed of 36 APIs, which is a fairly dramatic number. However, after 2012, the complexity of newly submitted mashups began to decline.

\textbf{The size of mashups in recent years has not increased}, which may imply that Web-based mashups cannot adapt well to the increasingly complex requirements of users. The popularity of the mobile Internet, the lack of open APIs, and the existence of commercial barriers may be other reasons for the decline in Web-based mashups.

\section{Threats to Validity}
In this section, we will discuss the threats to validity in detail. There are three main threats: \textbf{An Alternative Method}, \textbf{A More Appropriate Probability Distribution}, \textbf{Time Errors} and \textbf{Accuracy of Rule-based Techniques}.

\textbf{An Alternative Method}. There are two categories of methods to perform the estimation of death time: 
\begin{enumerate}
\item One is the overall estimation based on the probability distribution represented by the method proposed in Section~\ref{sec:estimate}. This kind of method is independent of the information in the specific API/mashup and can only guarantee a reasonable overall sample distribution but cannot provide a precise estimation.
\item Another method that we did not adopt is to construct a regression model and use it to predict the number of days each unavailable API/mashup survives. Ideally, this is the best method because it ensures both a reasonable overall distribution and individual accuracy. Unfortunately, this method is not practicable on the \textit{ProgrammableWeb} dataset. First, the amount of data is too small, which hinders deep learning-based regression models. Second, the information provided in the ProgrammableWeb dataset makes it difficult to identify features related to survival days, which hinders the feature engineering-based machine learning regression models.
\end{enumerate}
As we stated at the beginning of Section~\ref{sec:corrected}, our proposed method is not perfect and can only guarantee a reasonable overall data distribution, but this is sufficient in the scenario of evolutionary analysis. When we have more labeled data and richer information, it will be better to construct a regression model, which will produce more accurate results. Therefore, it is important to emphasize that the method proposed in this paper is a relatively reasonable and feasible way to address the evolutionary analysis scenario under the current data conditions.

\textbf{A More Appropriate Probability Distribution}. In Section~\ref{sec:estimate}, we assumed that the longevity of APIs/mashups follows a normal distribution and confirmed that the overall distribution and the data used for estimation were identically distributed by the Z-test. However, this does not indicate that the normal distribution is the best distribution. As more labeled data are provided, we may find that the survival days of the service follow a more complex distribution so that we can replace the normal distribution now used for estimation with a more accurate probability distribution. It should be noted that under the condition of limited observation data, it is not suitable to choose a complex distribution---it is more robust to utilize a simple distribution such as the normal distribution.

\textbf{Time Errors}. In Section~\ref{sec:estimate}, we select 20 APIs to check their dead times and found an error of 50 days. This error may slightly affect some of the descriptions in some of the research questions; for example, the turning points in RQ1 may be offset, but these do not affect the final conclusions of the research questions in this paper. We have also repeated the experiment several times to mitigate this threat.

\textbf{Rule-based Techniques}. We use rule-based techniques when checking the availability of APIs and Mashups. Due to the limitation of rule-based techniques, there may be a small number of false positive and false negative samples, which may result in slight variations in specific values, such as those in Table 1. To deal with this issue, we have taken repeated tests and manual sampling to verify the validity of these techniques. More manual checking and improving the rules-based technology can further reduce the risks associated with the rules-based technology.

\section{Conclusion}
\label{sec:conclusion}
In this paper, we analyzed the evolution of the \textit{Programm\-ableWeb} service ecosystem from a dynamic network perspective. We first summarized the quality issues in the original \textit{ProgrammableWeb} dataset and analyzed the negative effects of these quality issues on traditional service computing tasks. Then, we proposed novel methods to correct the evolution-related data quality issues, including API/mashup availability status, API/mashup death time, and mashup composition. Finally, we conducted a set of extensive experimental analyses on the corrected \textit{ProgrammableWeb} dataset, and the most intriguing finding is that the development of the \textit{ProgrammableWeb} service ecosystem is less optimistic than that reported by the existing studies. In fact, we discovered considerable evidence that the \textit{ProgrammableWeb} is declining. Our empirical analysis has identified a number of research challenges and opportunities. To ensure that the \textit{ProgrammableWeb} community continues to flourish, we encourage researchers, developers, and managers to address these challenges with innovative solutions.

\section*{Acknowledgment}
The research in this paper is partially supported by the National Key Research and Development Program of China (No 2018YFB1402500) and the National Science Foundation of China (61772155, 61832004, 61802089, 61832014).

\bibliographystyle{elsarticle-num}

\bibliography{cas-refs}

\end{document}